\documentclass[structabstract]{aa}  
\usepackage{graphicx}
\usepackage{epstopdf}
\usepackage{natbib}
\bibpunct{(}{)}{;}{a}{}{,} 
\usepackage{lscape}
\usepackage{longtable}
\usepackage{lscape}
\usepackage{color}
\usepackage{rotating}
\usepackage[percent]{overpic}
\usepackage{txfonts}

\usepackage{amstext}

\begin{document}

\title{New dwarf galaxy candidates in the Centaurus group}

   \author{Oliver M\"uller\inst{1}
          \and
          Helmut Jerjen\inst{2}
                                        \and
          Bruno Binggeli\inst{1}
          }

   \institute{Departement Physik, Universit\"at Basel, Klingelbergstr. 82, CH-4056 Basel, Switzerland\\
         \email{oliver89.mueller@unibas.ch; bruno.binggeli@unibas.ch}
         \and
         Research School of Astronomy and Astrophysics, The Australian National University, Mt Stromlo Observatory, via Cotter Rd, 
Weston, ACT 2611, Australia\\
         \email{helmut.jerjen@anu.edu.au}
             }

   \date{Received TBD; accepted TBD}

 
  \abstract
  {Recent studies of the distribution and kinematics of the Milky Way and Andromeda satellite galaxy systems have confirmed the existence of 
   coplanar, corotating structures of galaxies. In addition to the 'missing satellite problem', these structures pose a major challenge to the standard 
   $\Lambda$CDM scenario of structure formation.}
   {We complement the efforts made by the dwarf galaxy community to extend these studies to other nearby galaxy groups by systematically searching for faint{ unresolved} dwarf members with a low
surface brightness in the Southern Centaurus group of galaxies.
The aim is to determine whether these coplanar, corotating structures
are a universal phenomenon.}
   {We imaged an area of 60 square degrees (0.3\,Mpc$^2$) around the M\,83 subgroup with the wide-field Dark Energy Camera (DECam) at the 
   CTIO 4 m Blanco telescope in $g$ and $r$ down to a limiting surface brightness of $\mu_r\approx 30$\,mag\,arcsec$^{-2}$. Various image-filtering techniques were
   applied to the DECam data to enhance the visibility of extremely low-surface brightness objects.}
   {We report the discovery of 16 new dwarf galaxy candidates in the direction of the M\,83 subgroup, roughly doubling the number of known dwarfs in that region. The photometric properties of the candidates, when compared to those of the Local Group, suggest membership in the M\,83 subgroup. The faintest objects have a central star density of  $\approx1.3\,L_\odot$\,pc$^{-2}$ and a total magnitude of $g = 20.25$, corresponding to $M_g = -9.55$ at the nominal distance of 4.9\,Mpc. The sky distribution of the new objects is significantly prolonged toward Cen A, suggesting that many of them belong to the Cen\,A subgroup or a common halo. We also provide updated surface photometry for the brighter, known dwarf members in the surveyed area.}
 {Modern survey CCD cameras and sophisticated detection algorithms can be used to systematically probe the faint end of the galaxy luminosity function around the M\,83 subgroup of galaxies. We aim at finding more and fainter members over a larger area to obtain a complete picture of the satellite galaxy substructure in the Centaurus group { down to a total magnitude limit of $M_V \approx -10$}.}


   \maketitle
%

\section{Introduction}
Dwarf galaxies are key objects for the study of structure formation on galaxy scales. In particular, dwarf spheroidals are traditionally known to host large amounts of dark matter (DM) as inferred by their internal dynamics \citep[e.g.,][]{2007ApJ...670..313S, 2013pss5.book.1039W}. The mere abundance and large-scale distribution of these systems are an important testbed for DM and structure formation. There is the heavily discussed  so-called missing satellite problem, which describes the observed deficiency of dwarf satellites around the Milky Way galaxy when compared to the CDM prediction of DM subhalo numbers \citep{1999ApJ...524L..19M, 1999ApJ...522...82K, 2006ApJ...649....1D, 2008Natur.454..735D}. This discrepancy has not been substantially diminished by the discovery of numerous { ultra-faint} Milky Way satellite galaxies in the Sloan Digital Sky Survey
(SDSS) and other surveys \citep[e.g.,][and references therein]{2005ApJ...626L..85W, 2015ApJ...805..130K}. 
An even stronger challenge to structure formation with DM is posed by the highly asymmetric features in the distribution of dwarf satellites around the best-studied Milky Way and Andromeda galaxies. The Vast Polar Structure (VPOS) of the Milky Way \citep{2012MNRAS.423.1109P, 2015arXiv150507465P} and the Great Plane of Andromeda \citep[GPoA, ][]{2013Natur.493...62I} are planar structures that are difficult to accommodate in a standard  $\Lambda$CDM scenario \citep[e.g.,][ and references therein]{2014MNRAS.442.2362P}.

In the context of the existing disagreement between near-field cosmology observations and the predictions of the best cosmological models on galaxy scales, it is imperative to extend such studies on the abundance and distribution of dwarf galaxies to other nearby galaxy aggregates beyond the Local Group (LG), to see whether or not the relative sparseness and asymmetric distribution of faint dwarf satellite galaxies is a common phenomenon in the nearby Universe. Given the extreme 
surface brightness regime of the target galaxies and the large angular extents of nearby galaxy groups, this poses an observational challenge. { There is of course no hope, for many years to come, that we will be able to map the analogs of the ultra-faint ($M_V \geq -8$) dwarfs swarming around the Milky Way at larger distances. But present-day technology allows us to aim at a dwarf galaxy census of other nearby groups down to $M_V \approx -10$, cleary surpassing the achievements of the photographic Schmidt surveys and the  SDSS. Very recently, Tully et al.\,(2015) have reported hints of a double-planar structure in the Centarus group of galaxies, based on previously known (i.e., still fairly bright) dwarf members of the group. This is encouraging because it means that a study of the distribution of faint, but not necessarily ultra-faint, dwarfs in the nearby groups will provide important constraints for structure formation models.}

Several teams have taken up the effort to conduct dedicated { deep, wide-field} imaging surveys of nearby galaxy groups in the local volume closer than 10\,Mpc \citep{2005AJ....129..178K, 2013AJ....145..101K} to detect ever fainter dwarf galaxy members. 
In the northern hemisphere this was successfully done by \cite{2009AJ....137.3009C, 2013AJ....146..126C} for the M\,81 group and by \cite{2014ApJ...787L..37M} for the M\,101 group. 
In the same spirit, we have started a survey of the Centaurus and Sculptor groups in the southern hemisphere, based on images taken with the Dark Energy Camera (DECam) at the 4m Blanco telescope at Cerro Tololo Inter-American Observatory (CTIO) and the SkyMapper telescope of the Australian National University. A similar imaging survey of the two groups, albeit much more restricted in sky area but to greater depths, is underway \citep{2014ApJ...793L...7S}.

In this paper we report on our first harvest of candidate dwarf members in a region of 60 square degrees of the Centaurus group around M\,83. The Centaurus group is the richest and most massive group of galaxies in the local volume at a distance of $\approx$\,4 Mpc \citep[][Fig.\,2 therein, and following references]{1979AN....300..181K, 2000AJ....119..593J}, covering a huge area of $\approx$ 500 square degrees of southern sky. It can be considered the southern analog to the similarly rich and nearby M\,81 group in the northern hemisphere. 
As a result of various studies of the region conducted in the late 1990s that aimed to find new group members in the optical \citep{1997AJ....114.1313C, 1998A&AS..127..409K, 2000AJ....119..593J} and at 21\,cm \citep{1999ApJ...524..612B}, and the subsequent membership confirmation by distance measurements \citep{2000AJ....119..166J, 2002A&A...385...21K}, there are around 60 Centaurus group members known to date \citep[see][]{2002A&A...385...21K, 2014AJ....147...13K}. 
We note that the optical surveys mentioned above were all based on the same Schmidt photographic plates; an SDSS analog for the southern hemisphere with the SkyMapper telescope \citep{2007PASA...24....1K} has only just started. 
The Centaurus group has a pronounced bimodal structure, consisting of a larger aggregate centered on the massive peculiar galaxy NGC\,5128 = Centaurus\,A at a mean distance of 3.8 Mpc, and a smaller galaxy concentration around the giant spiral NGC\,5236 = M\,83 at a mean distance of 4.9 Mpc \citep{2002A&A...385...21K, 2014AJ....147...13K, 2015ApJ...802L..25T, 2015AJ....149..171T}. In the following, we refer to the main concentrations as the Cen\,A and M\,83 subgroup, respectively, the whole complex as the Centaurus group, or also Centaurus\,A group, as it is known in the literature up to the present.

The paper is organized as follows. In Sect. 2 we give the details of the imaging with DECam. Section 3 describes the techniques and the results of our search for faint diffuse dwarfs on the observed fields. In Sect. 4 we report on the photometry done for the new candidates as well as the known Cen members in the search area. Finally, a critical discussion of our findings is given in Sect. 5, followed by our concluding remarks in Sect. 6.

\section{Imaging}
Observations were conducted on 2014 July 17-19 using the Dark Energy Camera (DECam) at the CTIO 4m Blanco telescope as part of observing
proposal 2014A-0624 (PI: H. Jerjen). DECam has an array of sixty-two 2k $\times$ 4k CCD detectors with a 3 sqr deg field of view and a pixel scale of 0.27$\arcsec$ per pixel (unbinned). We obtained a series of 3$\times$40\,s dithered exposures in $g$ and $r$ bands under photometric conditions for 22 pointings with slight overlap. The mean seeing was 1.0$\arcsec$ in both filters. The fully reduced and stacked images were produced by the DECam community pipeline \citep{2014ASPC..485..379V}. Figure 1 shows the surveyed area around M\,83 with the known galaxies highlighted with filled circles. The new dwarf galaxy candidates are shown as stars. 

{ For the photometric calibration, we regularly observed Stripe
82 of the SDSS throughout the three nights with 50 s single exposures 
in each band. To determine the photometric zero points and color terms, we matched the instrumental
magnitudes of stars with the Stripe 82 stellar catalog to a depth of 23 mag and fit the 
following equations:
$$m_g = m_{g,instr} + Z_g + c_{g} \cdot (m_{g,instr}-m_{r,instr}) - k_{g}X$$
$$m_r = m_{r,instr} + Z_p+ c_{r} \cdot (m_{g,instr} - m_{r,instr}) - k_{r}X,$$
where $Z_g$ and $Z_r$ are the photometric zero points, $c_g$ and $c_r$ are
the respective color terms, $k_g$ and $k_r$ are the first-order extinctions, and $X$ is the mean airmass.
The best-fitting parameters are $Z_g=29.41$, $Z_r=29.45$, $c_r=0.126$,
and $c_g=0.092$. The
most recent extinction values $k_r=0.10$ and $k_g=0.20$ for CTIO were
obtained from the Dark Energy Survey team, and the airmass $X$ was given for the respective exposure.
To allow for a direct comparison with Local Group dwarfs, we converted the $gr$ photometry into the $V$ band \citep{SloanConv}
using the transformation equation
$$V = g - 0.5784\cdot(g - r) - 0.0038\,\,\,. $$
This formula applies to $m$ and $\mu$.
For the comparison of the S\'ersic parameters with Virgo and LG dwarfs we had to convert them 
from $B$ into $r$ band \citep{SloanConv} assuming a color index of 0.6 \citep{2008AJ....135..380L}
$$r= B-0.6-0.3130\cdot0.6-0.2271\,\,\,.$$
}

\begin{figure}[h]
  \resizebox{\hsize}{!}{\includegraphics{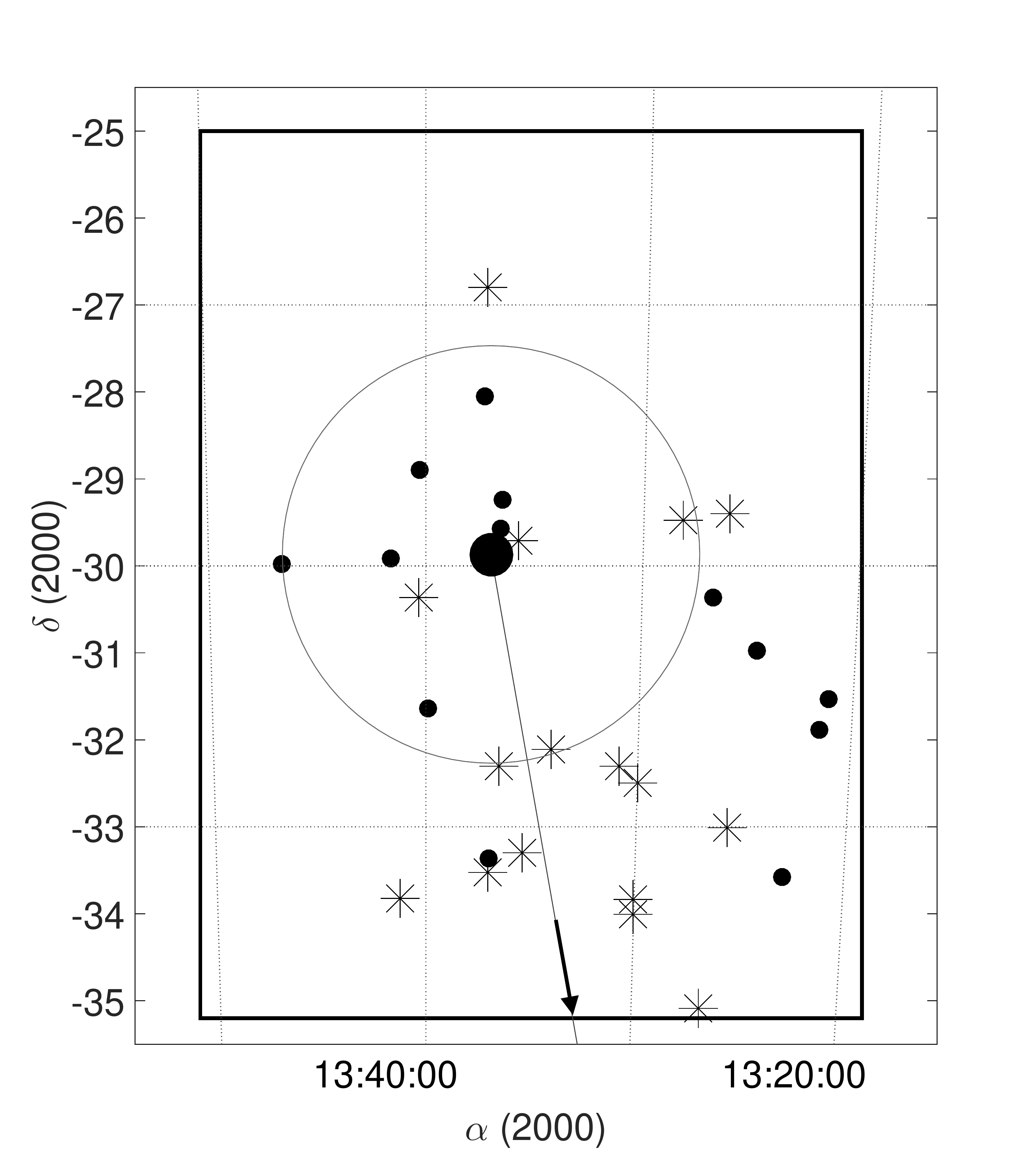}}
  \caption{Survey area of 60 sqr degrees (rectangle) around the galaxy M\,83 (large filled circle). The newly detected dwarf galaxy candidates are indicated with stars. 
Small filled circles are known M\,83 group members. The large circle indicates the virial radius of the M\,83 subgroup (see text).
The vector points toward the Cen\,A galaxy. 
We note an overdensity of new dwarfs in that direction.}
\end{figure}

\begin{table}
\caption{Names and coordinates (epoch 2000.0) of the new dwarf candidates.}
\label{table:1}
\begin{tabular}{lccl}
\hline\hline
Name & $\alpha$ & $\delta$ & Notes \\ \hline \\[-2mm]
dw1325-33 & 13:25:41 & $-$33:00:25 &  \\ 
dw1326-29 & 13:26:04 & $-$29:24:16 &  Irr?\\ 
dw1326-35 & 13:26:44 & $-$35:05:00 &  \\ 
dw1328-29 & 13:28:12 & $-$29:28:45 &  \\ 
dw1329-32 & 13:29:58 & $-$32:29:46 & BCD? \\ 
dw1330-32 & 13:30:54 & $-$32:18:21 &  \\ 
dw1330-33 & 13:30:04 & $-$33:50:06 &  Background?\\ 
dw1330-34 & 13:30:02 & $-$34:00:14 &  \\ 
dw1334-32 & 13:34:05 & $-$32:06:28 &  Cirrus?\\ 
dw1335-29 & 13:35:46 & $-$29:42:24 &  \\ 
dw1335-33 & 13:35:25 & $-$33:18:00 &  Cirrus?\\ 
dw1336-32 & 13:36:33 & $-$32:18:05 & \\ 
dw1337-26 & 13:37:13 & $-$26:48:10 &  \\ 
dw1337-33 & 13:37:02 & $-$33:31:25 &  \\ 
dw1340-30 & 13:40:19 & $-$30:21:35 &  \\ 
dw1341-33 & 13:41:13 & $-$33:49:30 &  \\ \hline
\end{tabular}
\end{table}

\section{Search and detection of new dwarf candidates}
With the chosen equipment and exposure time we do not expect any new dwarf spheroidal candidate at the distance of M\,83 to be resolved into stars. Taking the stellar population of the Sculptor dSph as a reference, we can assume $M_I = - 4.1$ and $V-I = 1.5$ for the TRGB of an old stellar population \citep{2007MNRAS.380.1255R}. With $V-r$ $\approx$ 0.2 this translates into $M_r$ $\approx$ $-2.8$, or at a distance of 4.9 Mpc and with $A_r = 0.15$ (see Table 2) into an expected $r$-band magnitude of 25.8 for the RGB tip. In comparison, the faintest stars on our DECam fields turned out to have $m_r$ between 25.0 and 25.5. Hence we clearly miss the tip, that is, we lack the stellar resolution of the suspected dwarf members of the M\,83 subgroup. The search for new dwarf candidates is therefore a search for extended unresolved
objects with a very low surface brightness in the fields.

Different filtering techniques and gray-level manipulations were applied to enhance the 
contrast and low surface brightness features on the images. Each of the reduced $g$ and $r$ frames were individually visually inspected and then stacked and inspected again. The dynamical range of the screen display was set so that the lowest and highest level was 2 sigma above and below the sky background. This provides the optimal resolution in the intensity regime where we expect the low surface brightness features of faint dwarf galaxies. The most important advantage of observing the $g$ and $r$ frames separately is that we can rule out artifacts such as noise fluctuations or scattered light. Finally, we used two different filtering techniques, a Gaussian convolution and the ring median filter \citep{1995PASP..107..496S}, to enhance the images even further. The ring median filter replaces each pixel value with the median of a narrow annulus around the pixel at a characteristic radius $r_c$. This filter has the effect of removing objects smaller than $r_c$ but leaving objects larger than $r_c$ mostly unaffected. Using these two filters in combination with different values for kernel width and radius revealed the same number of low surface brightness dwarf candidates. We note that the discovery rate by naked eye is surprisingly high in comparison with computer-based detections. \citet{2009AJ....137.3009C} and \citet{2014ApJ...787L..37M} drew similar conclusions. The list with the coordinates of the 16 new dwarf galaxy candidates is presented in Table 1. A gallery of the $r$-band images of the candidates is given in Fig.\,2.

\begin{figure*}
\includegraphics[width=4cm]{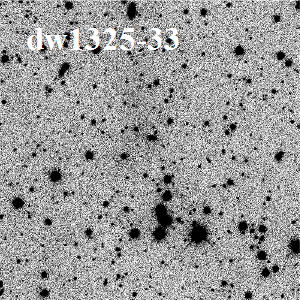}
\includegraphics[width=4cm]{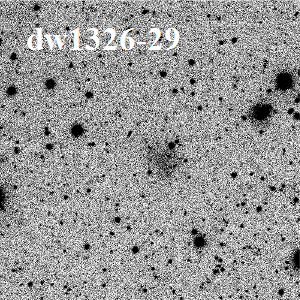}
\includegraphics[width=4cm]{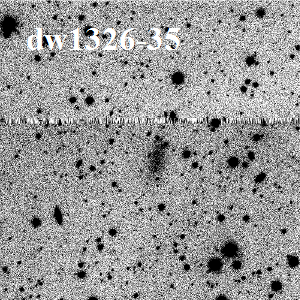}
\includegraphics[width=4cm]{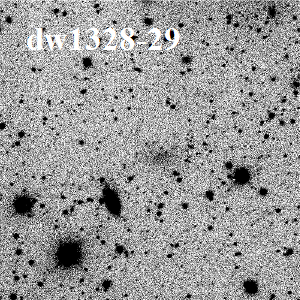} \\
\includegraphics[width=4cm]{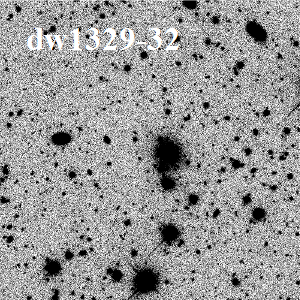}
\includegraphics[width=4cm]{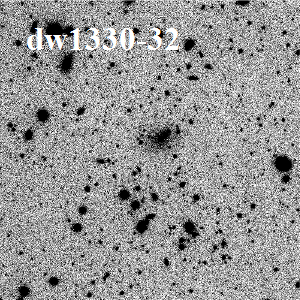}
\includegraphics[width=4cm]{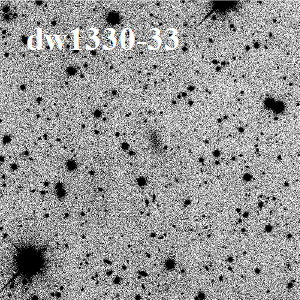}
\includegraphics[width=4cm]{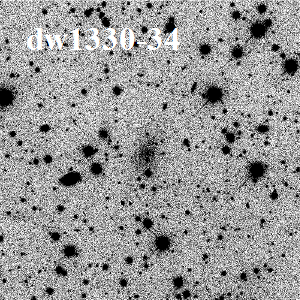} \\
\includegraphics[width=4cm]{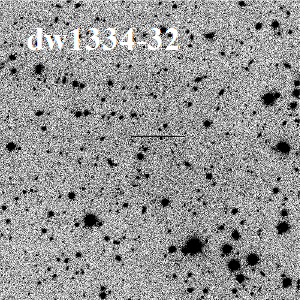}
\includegraphics[width=4cm]{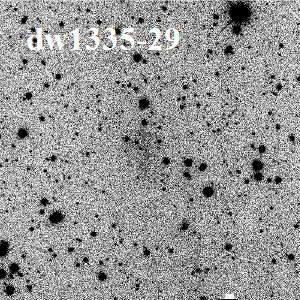}
\includegraphics[width=4cm]{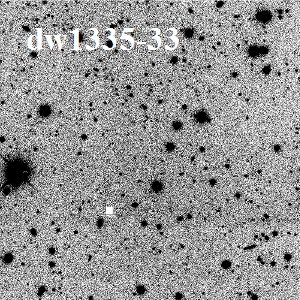}
\includegraphics[width=4cm]{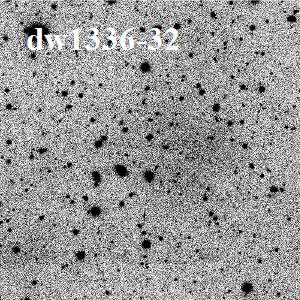} \\
\includegraphics[width=4cm]{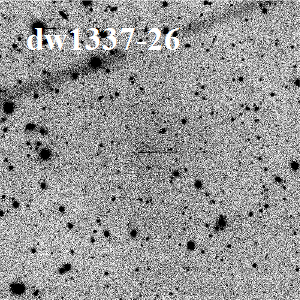}
\includegraphics[width=4cm]{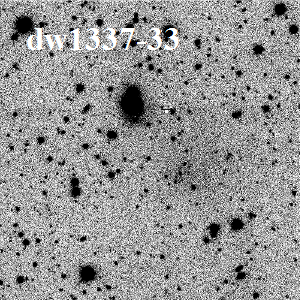}
\includegraphics[width=4cm]{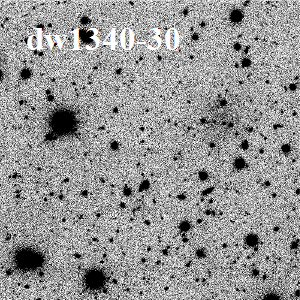}
\includegraphics[width=4cm]{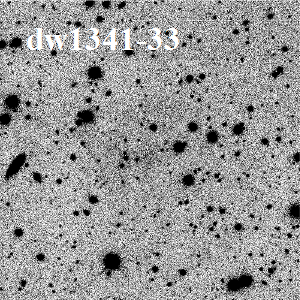} \\
\caption{Gallery of $r$-band images of all new M\,83 subgroup dwarf galaxy candidates. One side of an image is 4.5 arcmin or 6.6\,kpc at 4.9\,Mpc.}
\end{figure*}

We note that the relatively high Galactic foreground in the direction of the M\,83 subgroup could be a major obstacle when searching for new dwarf candidates. 
Therefore we reanalyzed the DECam images after removing foreground stars with the Otsu threshholding method \citep{Otsu} and 
with SourceExtractor \citep{1996A&AS..117..393B}. No more dwarf galaxy candidates have been detected with these methods.
Another difficulty of high Galactic foreground is the presence of very faint, low surface brightness $\text{cirrus}$ clouds (reflection nebulae), which can closely mimick faint dwarf galaxies. This is a known problem also in the region of the M\,81 group \citep{2009AJ....137.3009C}. Most cirrus is, however, morphologically distinct due to its large angular size, flat SB profile, or the presence of sharp edges. Still, in two cases (candidates dw1334-32 and dw1335-33, see Table 1) we cannot entirely exclude the possibility that we were misled in this way. The two objects are huge (Fig.2), very flat (Fig.5), and one of them (dw1334-32) is also unusually blue (Table 2).

{ To assess the efficiency and brightness limitations of our detection methods, we blindly added a number of artificial galaxies to several fields at random positions. Assuming exponential profiles (i.e., S\'ersic $n$ = 1, which is roughly fulfilled by our candidates, see below), the $r$-band total magnitude and central surface brighntess of the fake galaxies were varied 
from $m_r$ = 17 to 20 mag and from $\mu_{0,r}$ = 25 to 18 mag arcsec$^{-2}$, respectively. In this way, judged a posteriori, the brightest galaxies would certainly be visible or detectable, the faintest galaxies certainly not. Varying the two parameters independently in half-magnitude steps led to a set of 7$\times$7 = 49 artificial galaxies that were ingested in one field (corresponding to one of the 22 pointings). As the whole region is densely covered with stars and occasionally also cirrus patches, which will render the detectablity of a faint dwarf difficult, the same set of artificial galaxies was randomly placed five times in a given field. If a test galaxy with a given magnitude and surface brightness was discovered by the same methods used for the real galaxies in one or more cases, it was judged as detectable in principle. The number of its detection at different random positions is then a measure of the detection efficiency. This whole procedure was applied for five fields (pointings) distributed over the whole search area (one central field and four fields toward the corners) to test for positional variations.

Figure\,3 gives an impression of an artificial galaxy as compared to a real one with the same photometric parameters. Figure\,4 gives the results of the whole test procedure. Shown are the 49 parameter combinations of the artificial galaxies in the $\mu-m$ plane with their efficiency values averaged over all five fields (0 = no detection, 10 = 100\% detection). As expected, the detection limit is primarily given by the surface brightness. Below a central $\mu$ of 27, our artificial galaxies could rarely be recovered (if so, then it was more likely due to a mismatch with sky background noise). $\mu_{0,r}$ = 27 mag arcsec$^{-2}$ is indeed also the surface brightness limit of our dwarf candidates (see Fig.\,6 and Table 2). The recovery efficiency is 80 to 90\% for bright galaxies (at the upper left in the $\mu-m$ plane), significantly dropping near the surface brightness limit (downward), but also, albeit to a lesser degree, with increasing compactness (steeper profile, toward the right in the $\mu-m$ plane). As the typical dwarf galaxy is discovered by its characteristic low surface brightness {\em and} flat profile, this is no surprise. Classically, a  completeness boundary for a galaxy survey is drawn on the assumption that all galaxies that are larger than a given diameter (= 2 $r_{lim}$) within a given isophotal level of $\mu_{lim}$ are detected. For exponential profiles this two-parametric completeness limit turns into the form \citep[see][]{1990PhDT.........1F, 1988AJ.....96.1520F}
$$m_{tot}= \mu_0 - 5\log{(r_{lim})} - 2.18 + 5\log{(\mu_{lim}-\mu_0)}$$
With $r_{lim}$ = 20 arcsecs and $\mu_{lim}$ = 28 mag arcsec$^{-2}$ (in the $r$ or $V$ band, there is only a $\approx$ 0.2 mag difference), this curve gives a fair description of the completeness boundary for our real dwarf candidates, as shown in Fig.\,6 below. The same curve is plotted here in Fig.\,4 for the artificial galaxies. Clearly, some of the more compact (steep profile) artificials recovered as possible dwarfs fall outside (to the right) of this completeness boundary. With one exception, no dwarf candidates in the corresponding parameter range were found, which is of course due to the $\mu-m$ relation for dwarf galaxies (see Fig.\,6). 

We finally note that {\em \textup{no}} positional variation of the detection efficiency was found. Differences among the five fields are within Poisson noise. This is indeed expected because there is no strong stellar density gradient over the face of the survey area.} 

We did not attempt to morphologically classify our dwarf candidates at this stage of first imaging. Most of them will, once confirmed, be dwarf spheroidals. Two candidates show signs of irregularity, as noted in Table 1. One of them is a possible blue compact dwarf (BCD, dw1329-32) according to a clear hint of resolution into (presumably) knots of active star formation. We note that our search for new dwarf members is clearly biased toward low surface brightness objects through
the adopted search method. There might be more dwarf members of a compact nature.

\begin{figure}
  \includegraphics[width=4.45cm]{dw1325-33_r.png}
        \includegraphics[width=4.45cm]{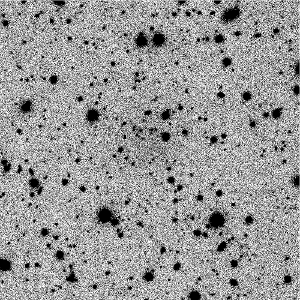}
\caption{{ Dwarf candidate dw1325-33 (left) and its artificial twin at a random position (right), easily detectable at a $\mu_{0,r} \approx$ 26 mag arcsec$^{-2}$ .}}
\end{figure}

\begin{figure}
  \includegraphics[width=8.5cm]{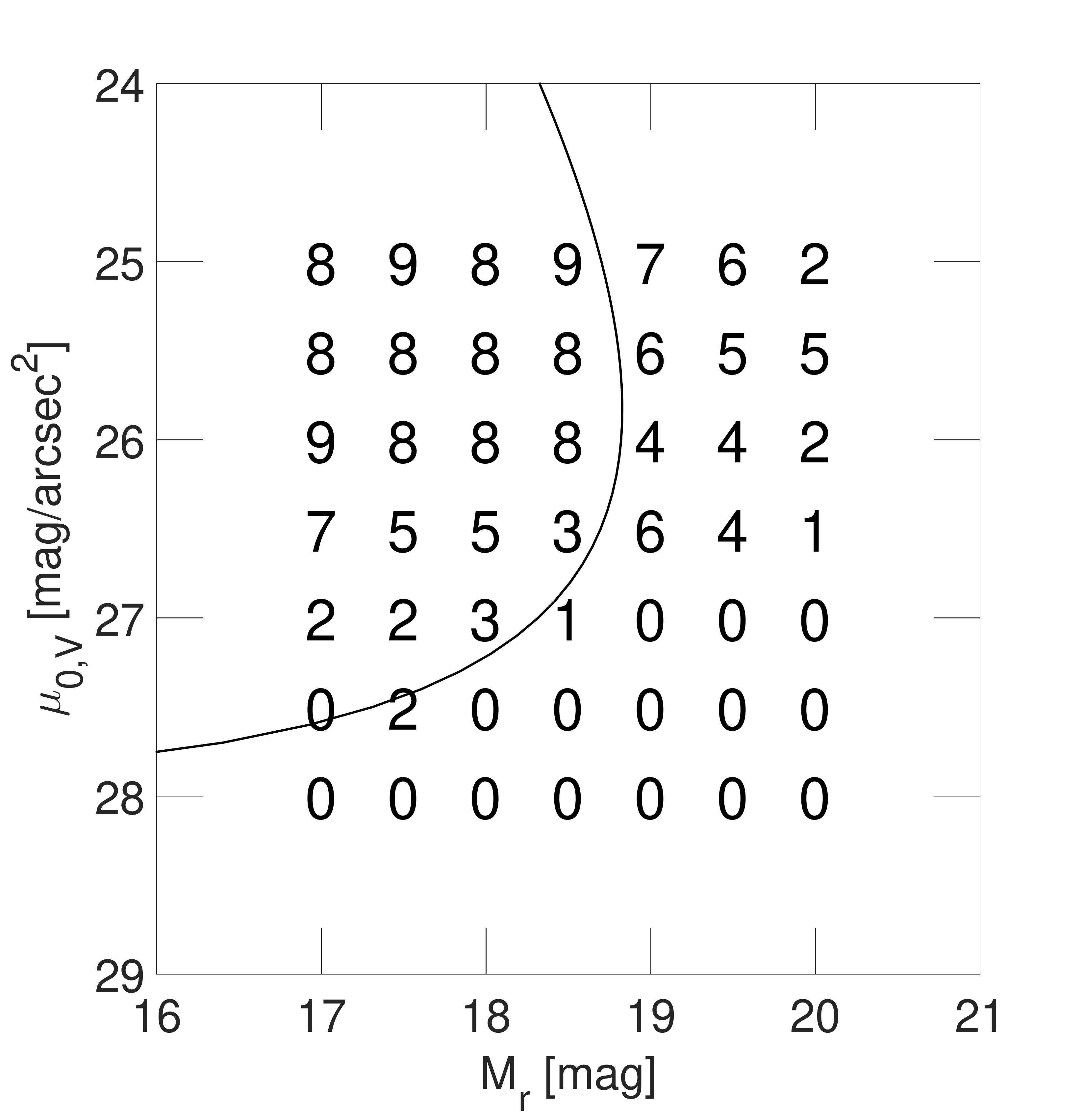}
\caption{{ Set of 49 artificial galaxies in the $\mu-m$ plane. The numbers indicate the detection efficiency (0 = no detection, 1...10 = 10\%...100\% detection). The curve gives a completeness boundary for the real dwarf candidates (see text and Fig.\,6).}}
\end{figure}


\section{Photometry}

Surface photometry was measured for all new dwarf galaxy candidates and for twelve of the thirteen known dwarfs in the survey area (KK208 \citep{1998A&AS..127..409K} was dismissed because it lacks a well-defined center). Great care was taken when removing foreground stars and background galaxies in the vicinity of the dwarf galaxy candidate by replacing the affected pixels with patches of sky from the surrounding area. The IRAF command $radprofil$ written by one of us was used to compute the total instrumental apparent magnitude $m_{instr}$, the effective radius $r_{eff}$, mean effective surface brightness $\langle \mu\rangle_{eff} $ , and the best-fitting S\'ersic \citep{1968adga.book.....S}  parameters $r_{0}$, $\mu_{0}$ , and $n$ in both $g$ and $r$ bands. The center of each object was estimated from the faint outer isophotes, which does not necessarily coincide with the peak surface brightness. The sky brightness was determined from the asymptotic behavior of the galaxy growth curve (cumulative intensity profile). We measured the radial surface brightness profile with a radial step size of 1.35 arcsec, except for dw1330-33, where a smaller step size of 0.54 arcsec was required given the small 
angular size of the galaxy. All profiles are shown in Fig.\,5. S\'ersic profiles were fitted at the surface brightness profiles using the equation
$$\mu_{sersic}= \mu_0+1.0857\cdot\left(\frac{r}{r_0}\right)^{n}$$
where $\mu_0$ is the S\'ersic central surface brightness, $r_0$ the S\'ersic scale length, and $n$ the S\'ersic curvature index.
We note that some authors use $1/n$ instead of $n$.

We give a general error estimate for magnitudes on the order of 0.3. Errors arise from the star subtraction ($\approx$ 0.2 mag, difference in magnitude between the galaxy with star removal and without), zero-point calibration (less than 0.04\,mag) and the determination of the sky from the variation of the asymptotic growth curve ($\approx$ 0.2 mag, given by the highest and lowest value of the sky level where the end of the growth curve is still asymptotically flat). For absolute magnitudes there are the additional errors based on the assumed distance (0.5 mag for an uncertainty in distance of $\pm0.5$ Mpc) and Galactic extinction (typically 0.015 mag). We assumed a mean distance for the M\,83 subgroup of 4.9 Mpc \citep{2002A&A...385...21K, 2014AJ....147...13K}. If some of the candidates are found to belong to the closer Cen\,A subgroup (see below), they would be intrinsically brighter by up to 0.7 mag.
Uncertainties for the half-light radius $\Delta r_{eff}=1.3$ arcsec and the mean effective surface brightness $\Delta \langle\mu\rangle_{eff}=0.01$ mag,{ not accounting for the magnitude uncertainty}, were again determined by the variation of the growth curve. The errors for the S\'ersic parameters are based on numerics and are given in Table 2. We note that the galaxy NGC\,5264 was located 
at the edge of the DECam image and thus has a higher level of uncertainty. 

{ Three of the already known dwarfs (CenA-dE2, Cen8, and CenA-dE4) are part of the sample of Jerjen et al.\,(2000a). A comparison of the photometric parameters shows good agreement; differences in magnitudes and surface brightness are within 0.3 mag. A systematic comparison of our photometry with other mostly much older literature sources is beyond the scope of this paper. To check the internal consistency of our photometry, we also applied our surface photometry machinery to a number of our artificial galaxies (Sect.\,3). The central surface brightness could always be reproduced to within 0.1 mag, the exponential scale to within a factor of 1.1. To reproduce the total magnitude to within less than 0.3 mag, we had to choose an extra-large aperture; real galaxies obviously have a cutoff and do not extend to infinity as the exponential model does. These artificial galaxy tests have confirmed the error estimates we reported above. 
The photometry we present, of the candidates and the known dwarfs alike, has the main purpose of assessing the plausibility of our candidates as dwarf members of the Centaurus group.}

In Table 2 we present the photometric data of all dwarf candidates and known dwarfs in the survey area. For KK208 we were unable to do photometry because it lacks a well-defined center. The items listed are as follows:
(1) Name of candidate or previously known dwarf. 
(2) Total apparent magnitude in the $r$ band. 
(3) Total apparent magnitude in the $g$ band. 
(4) Galactic extinction in $r$ according to \citet{2011ApJ...737..103S}.
(5) Galactic extinction in $g$ according to \citet{2011ApJ...737..103S}.
(6) Absolute magnitude in the $r$ band. The assumed distance for the candidates is 4.92 Mpc. For the known dwarfs TRGB distances were adopted according to \cite{2013AJ....145..101K}. CenA-dE2, CenA-dE4, and CEN8 have no TRGB distance, therefore the nominal distance of 4.92 Mpc is applied.
(7) Extinction-corrected color index $g-r$.
(8) S\'ersic central surface brightness in the $r$ band.
(9) S\'ersic scale length in the $r$ band.
(10) S\'ersic curvature index in the $r$ band.
(11) Mean effective surface brightness in the $r$ band.
(12) Effective radius in the $r$ band.

\begin{figure*}
\centering
\includegraphics[width=4cm]{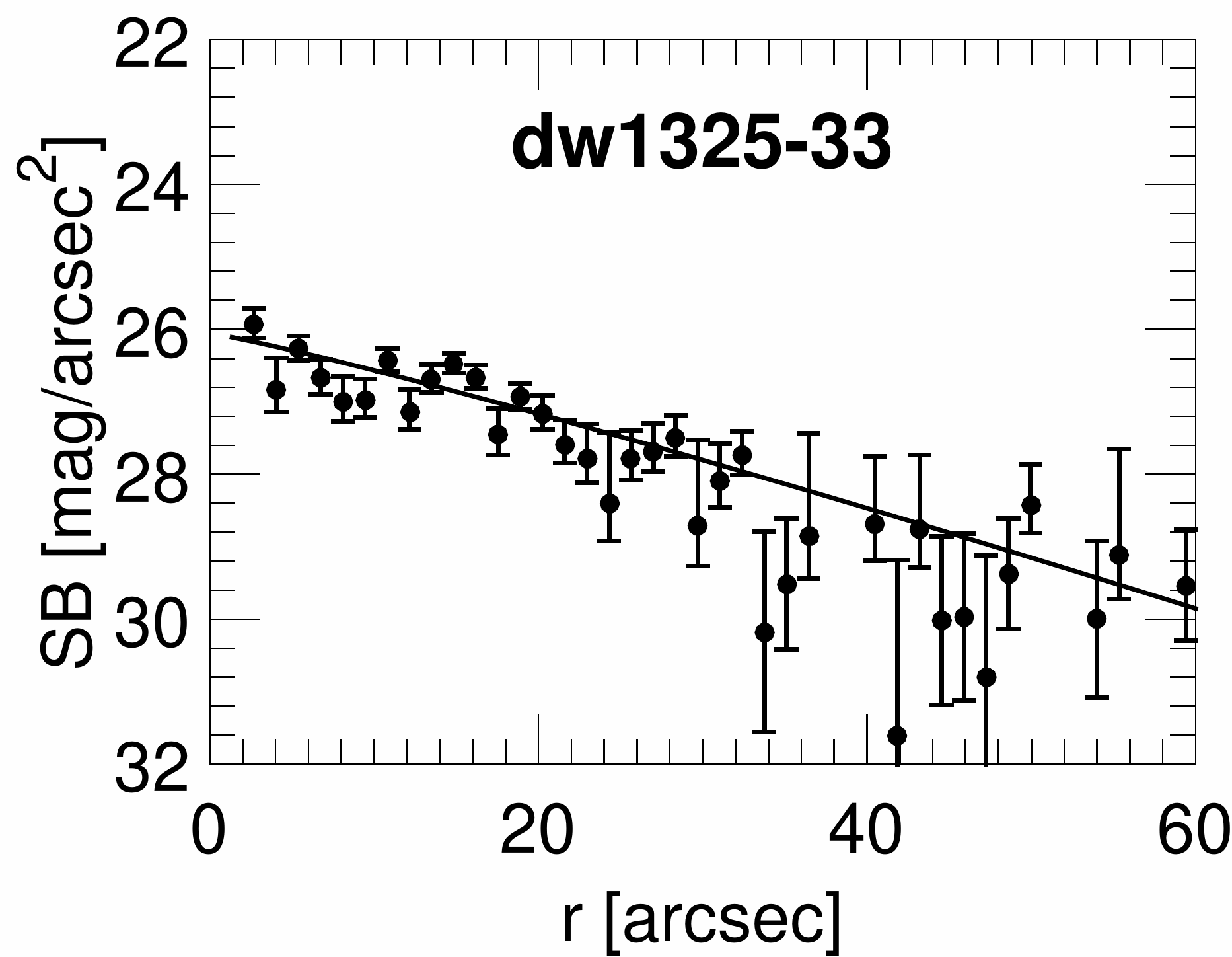}
\includegraphics[width=4cm]{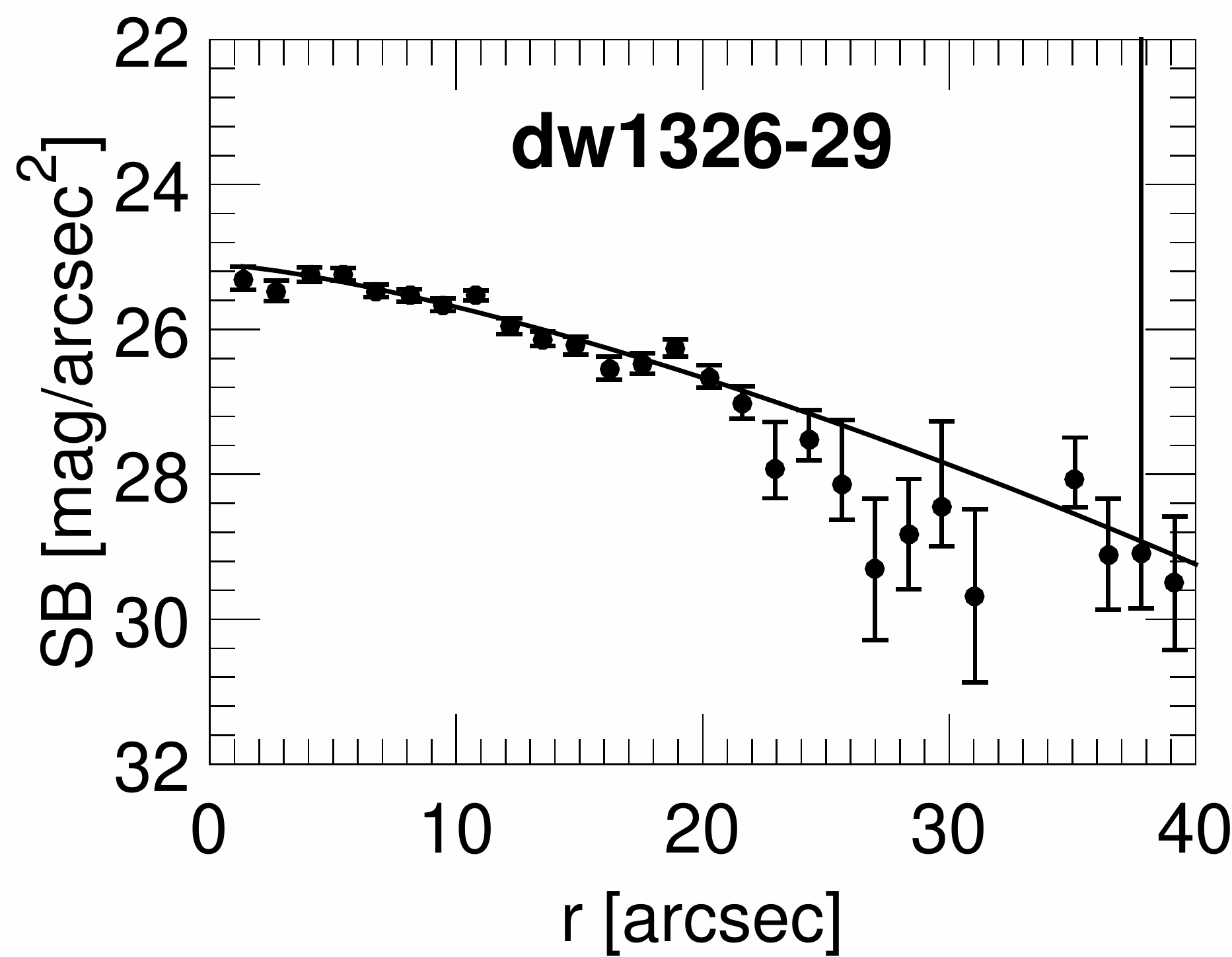}
\includegraphics[width=4cm]{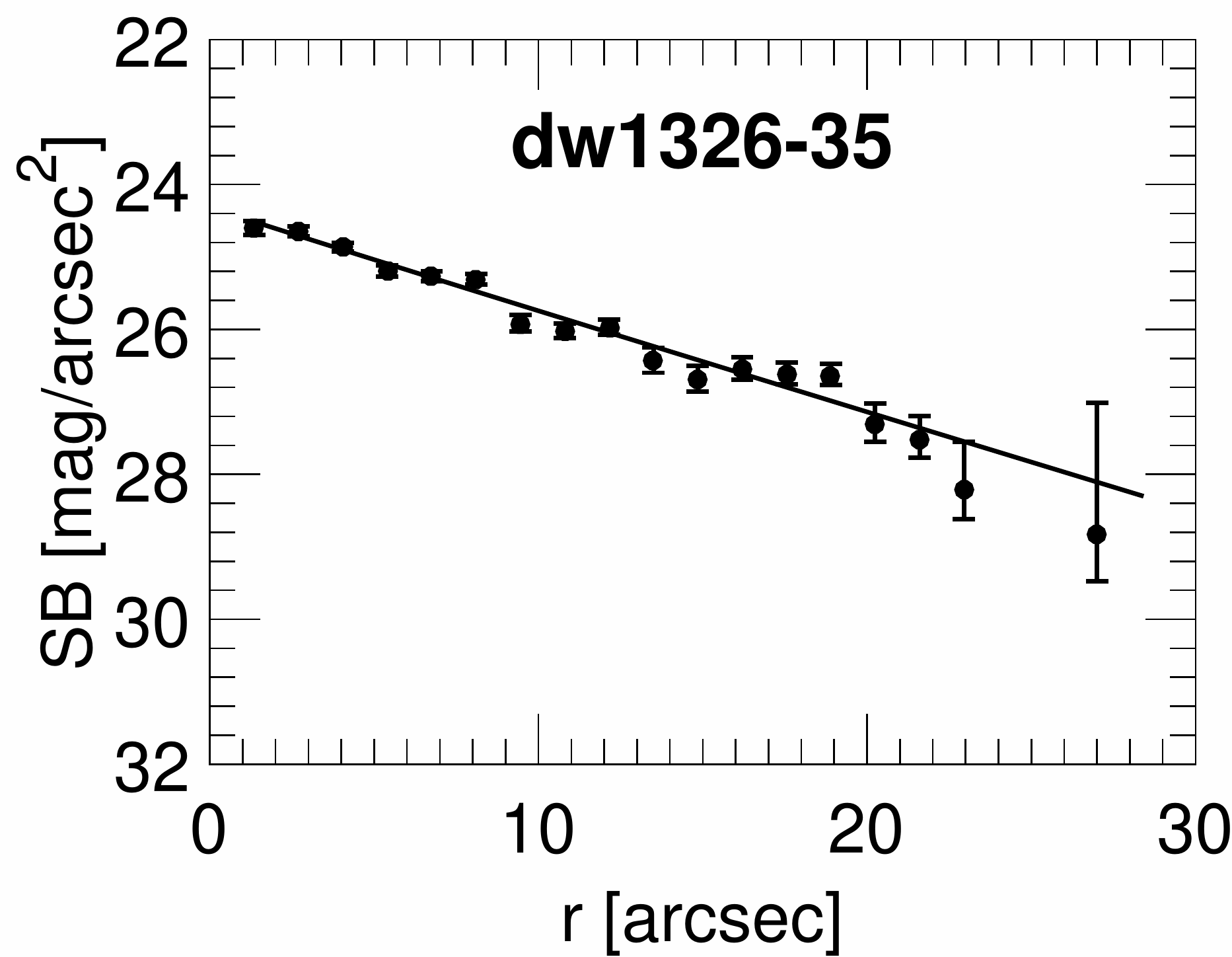}
\includegraphics[width=4cm]{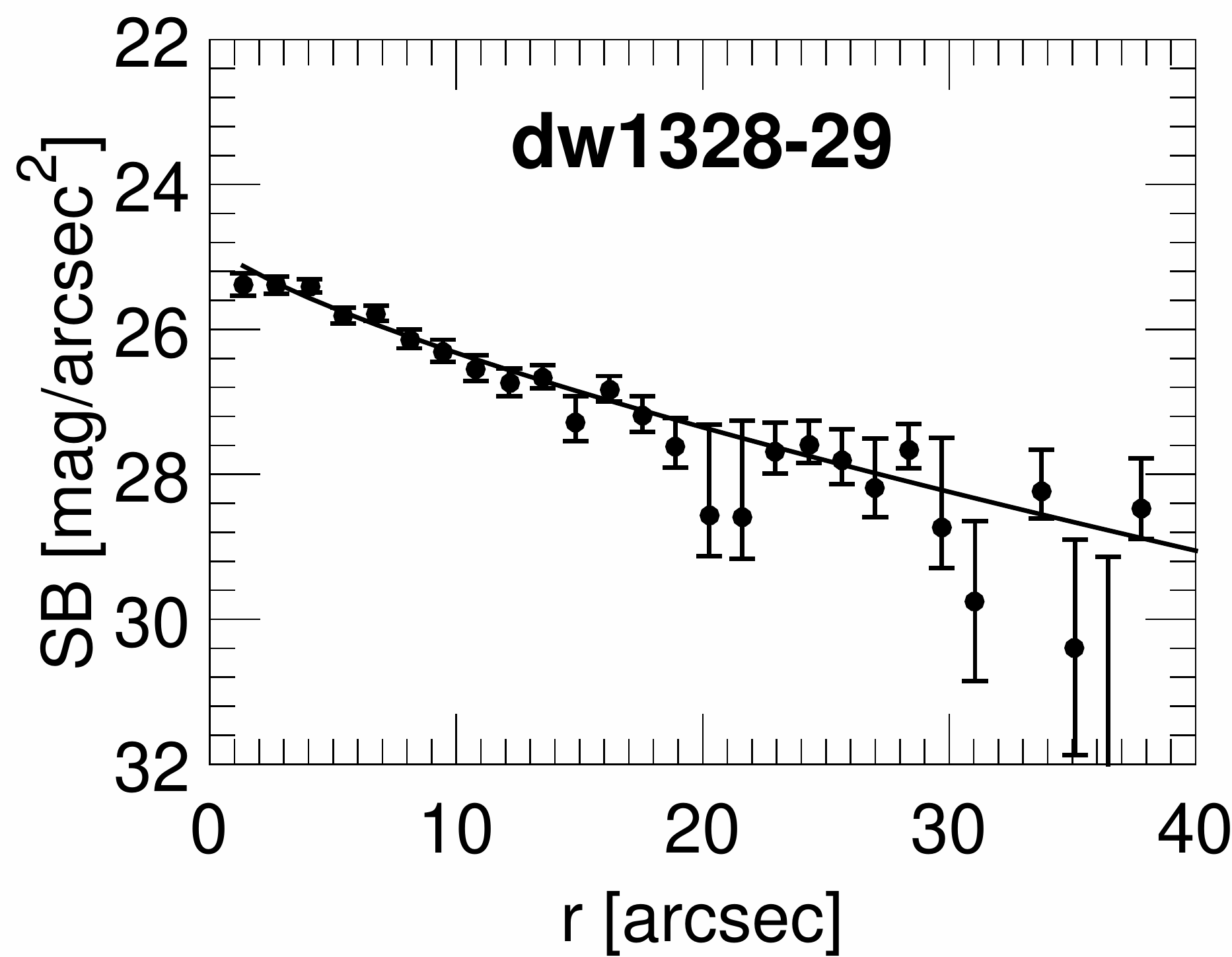} \\
\includegraphics[width=4cm]{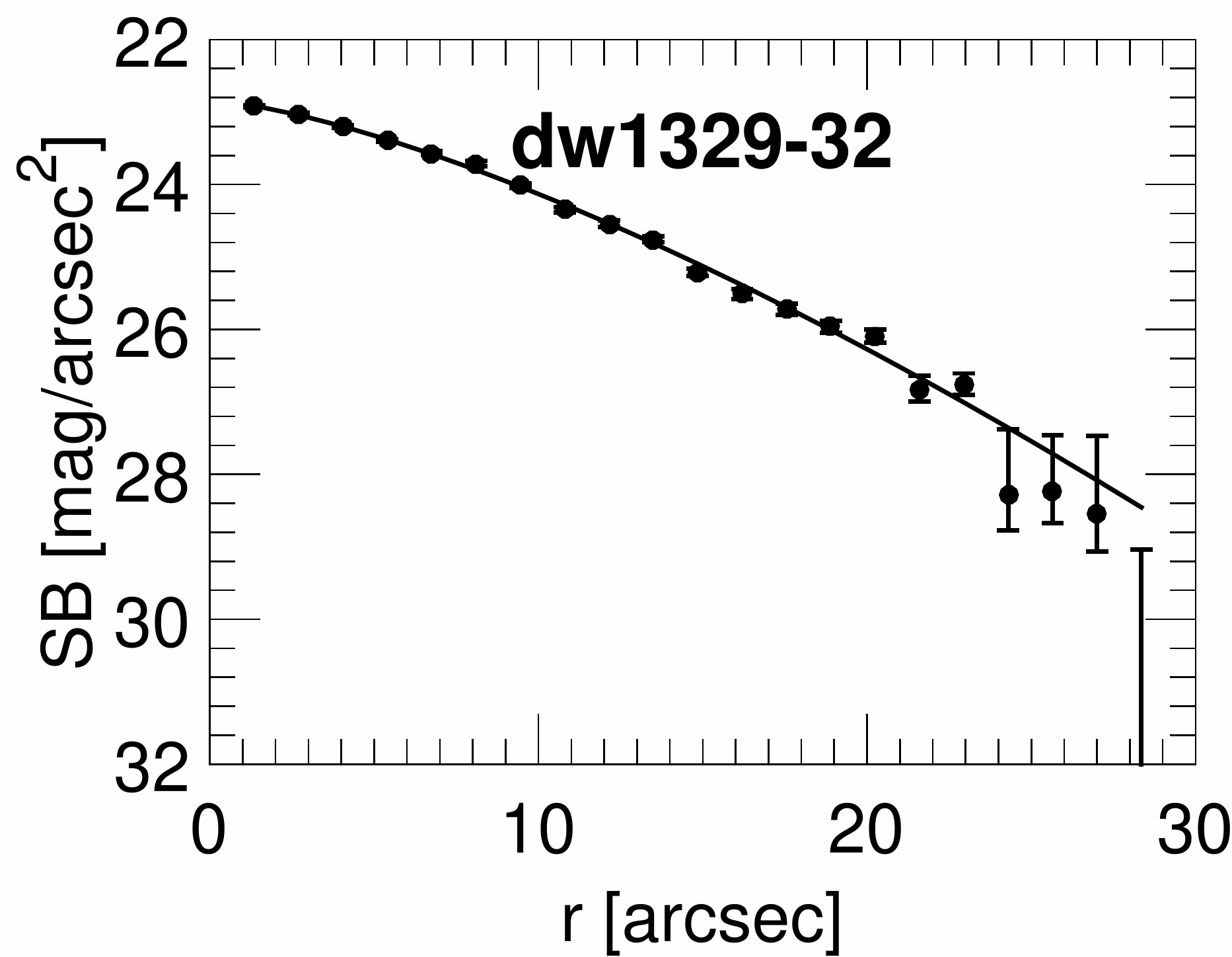}
\includegraphics[width=4cm]{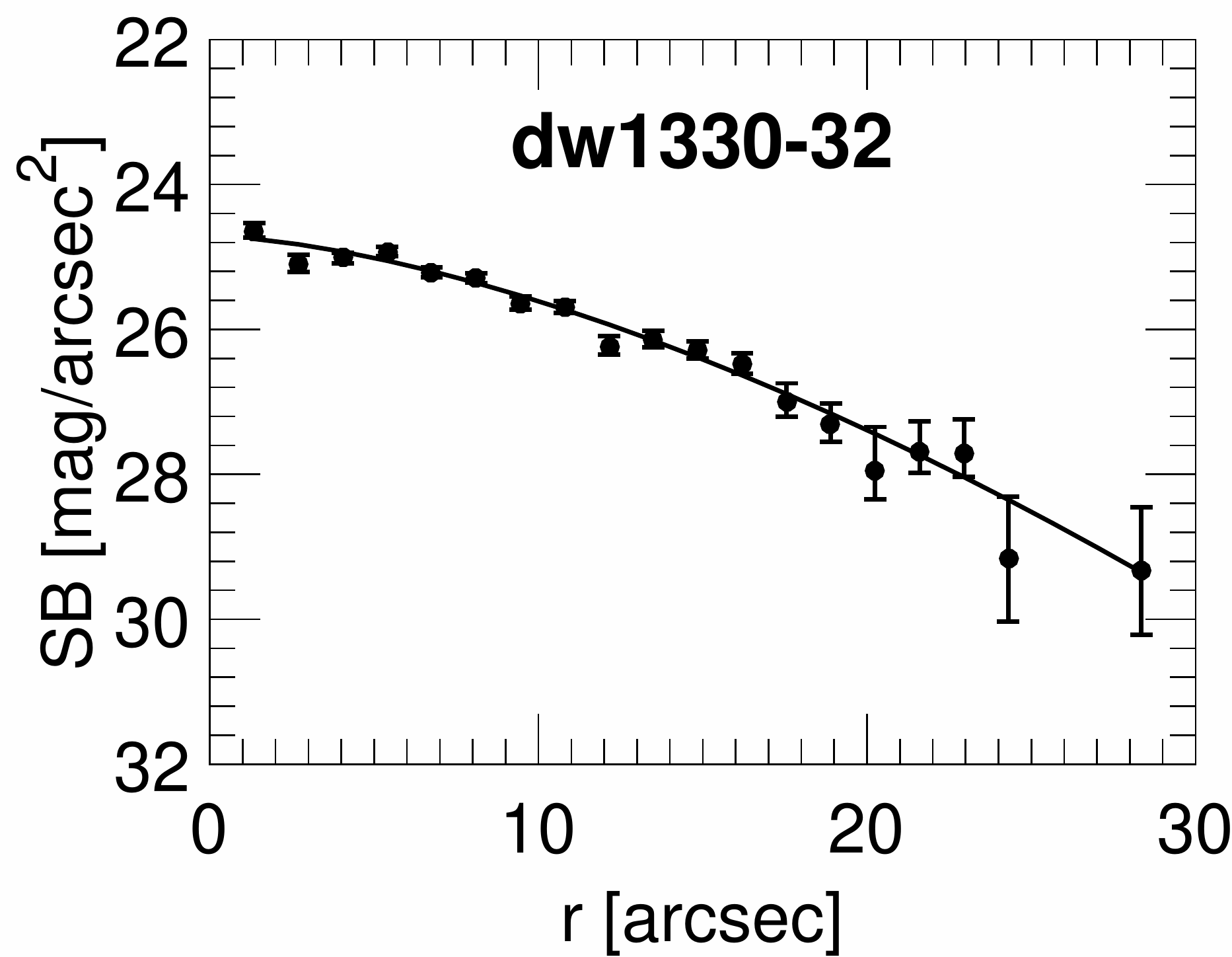}
\includegraphics[width=4cm]{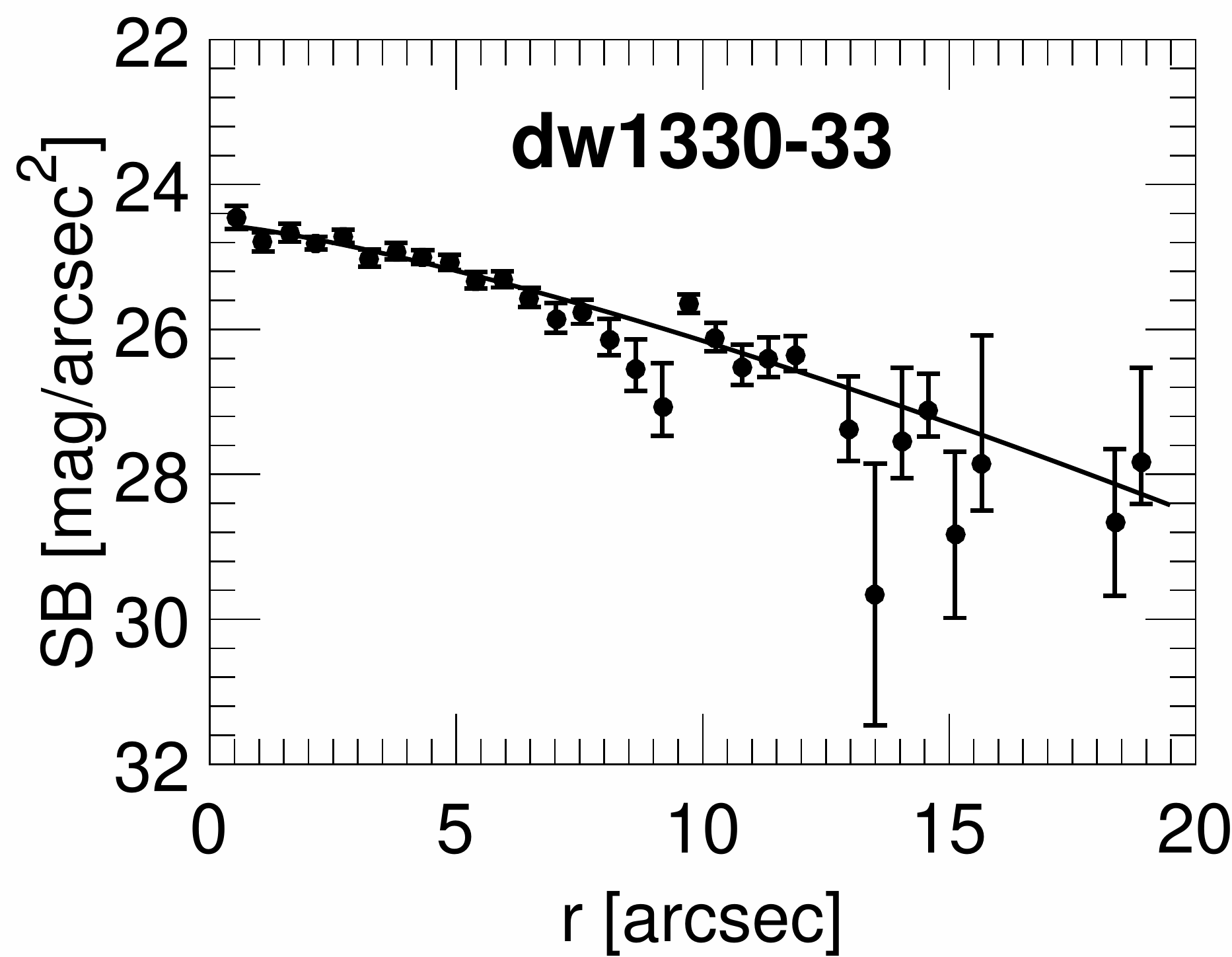}
\includegraphics[width=4cm]{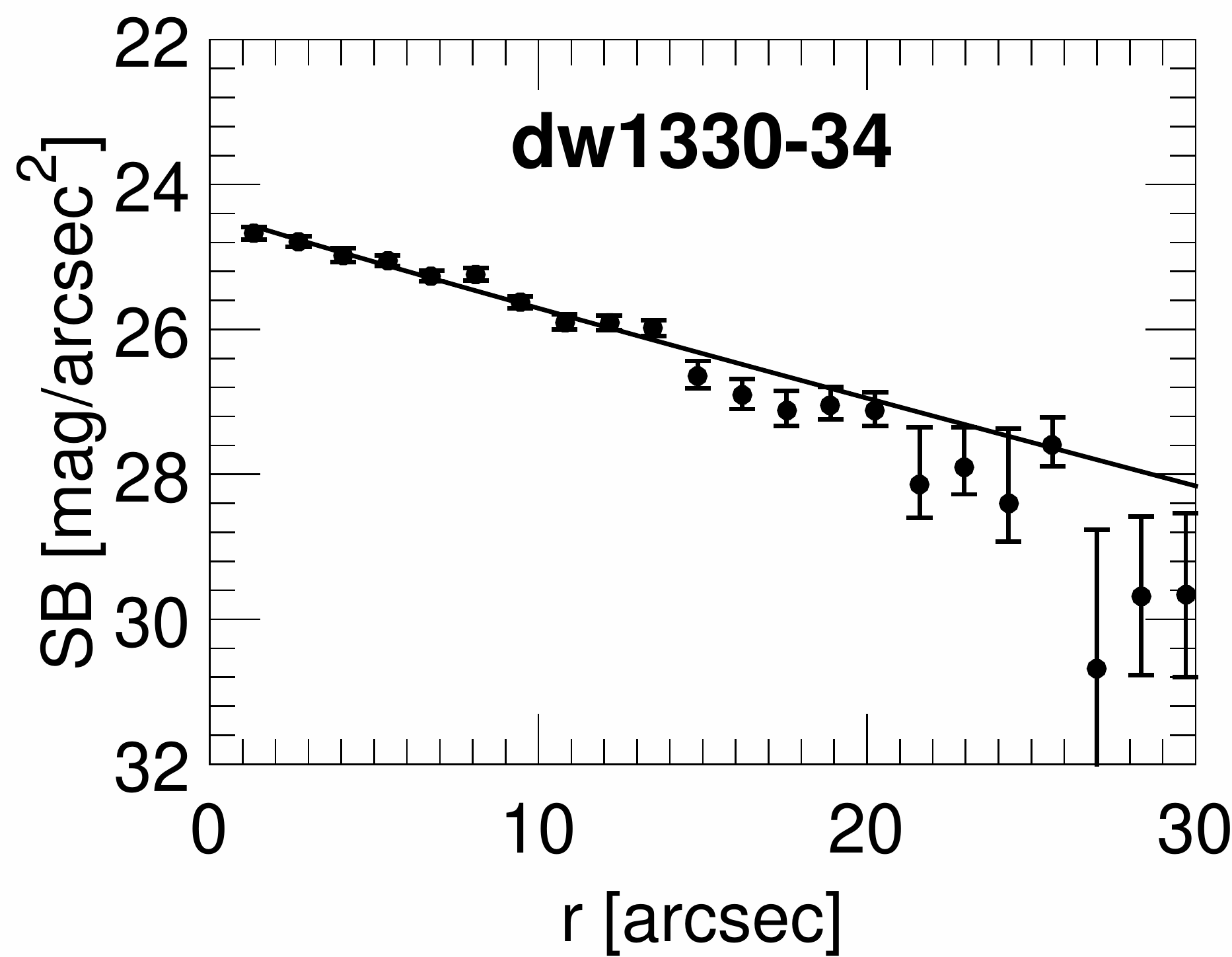} \\
\includegraphics[width=4cm]{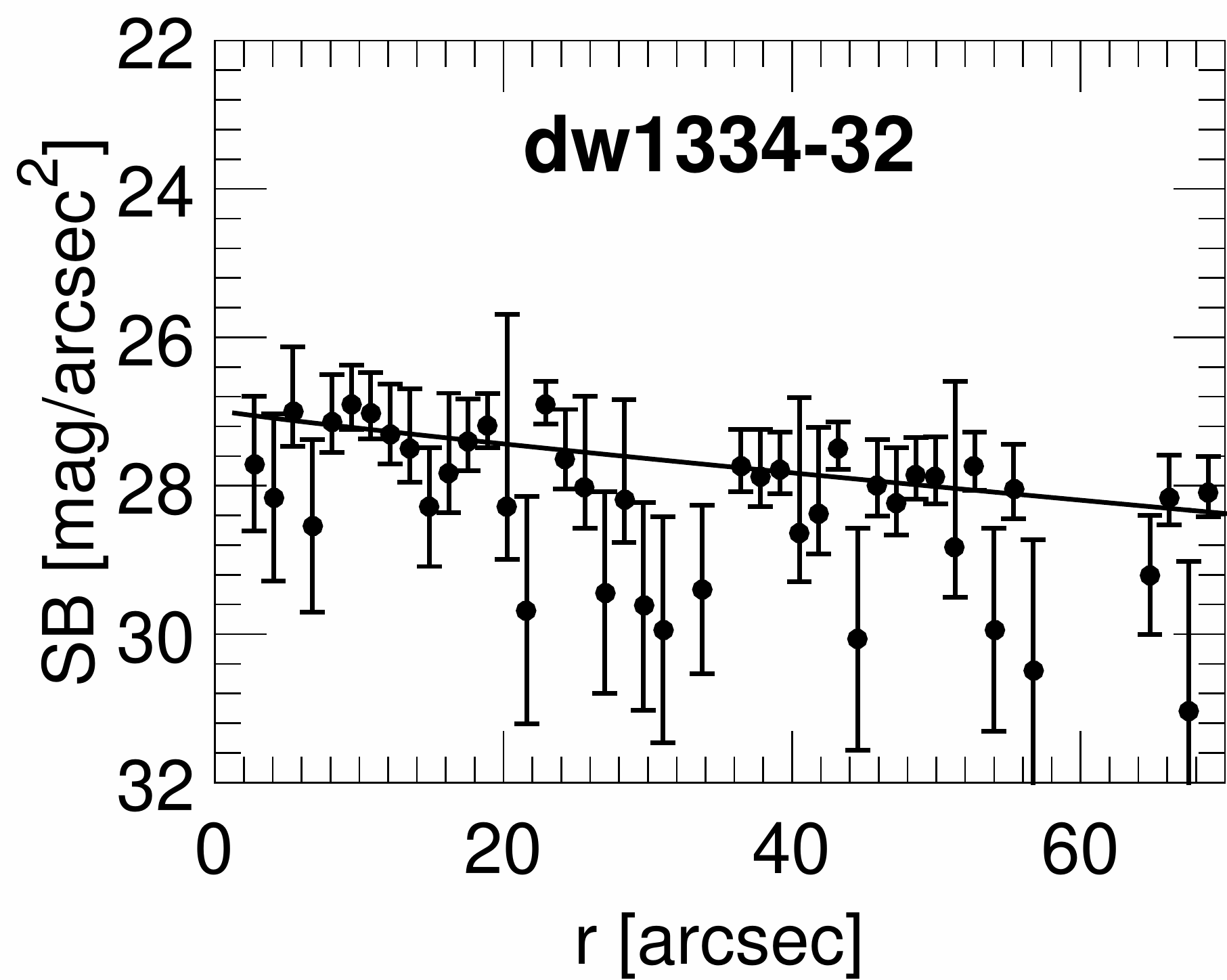}
\includegraphics[width=4cm]{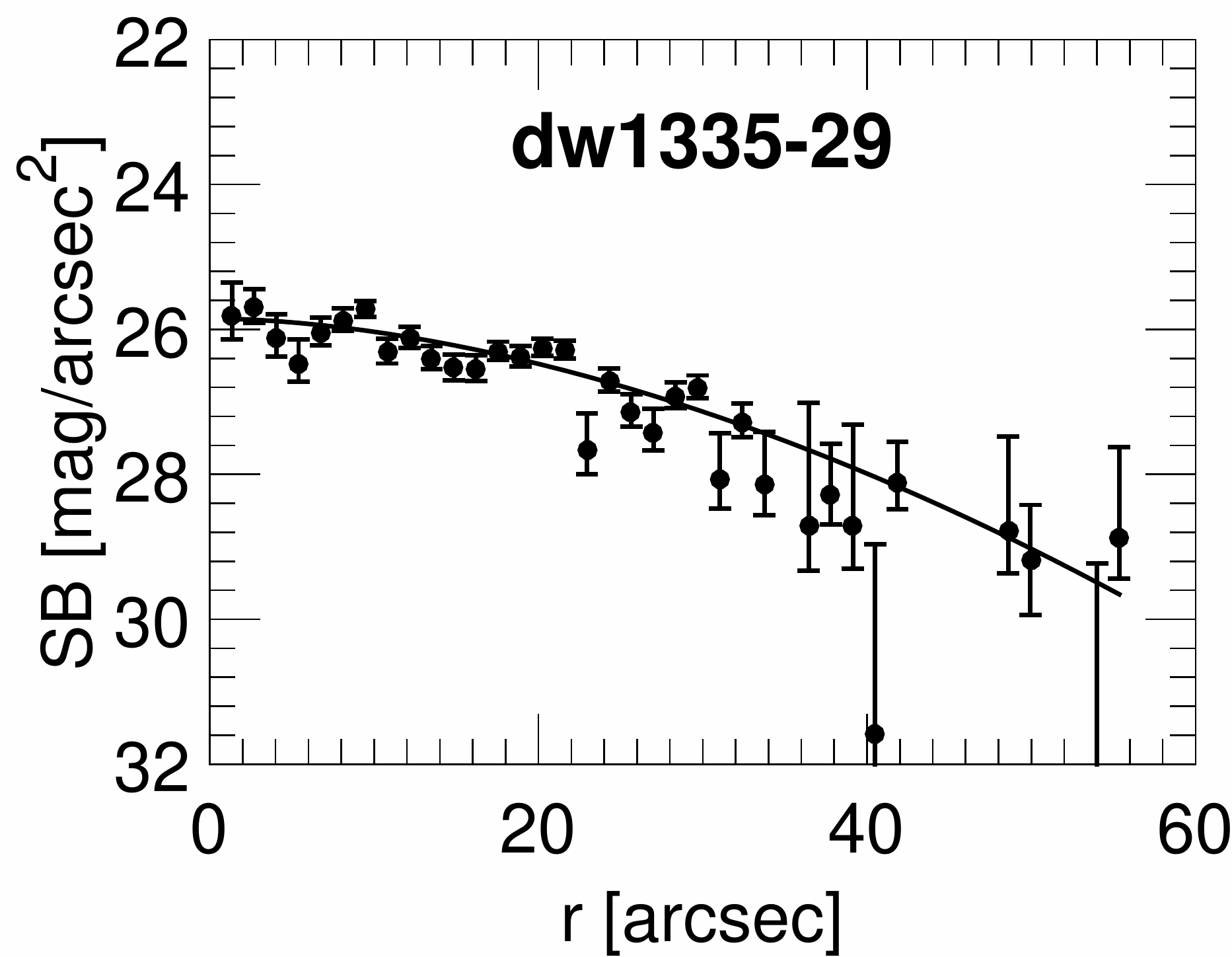}
\includegraphics[width=4cm]{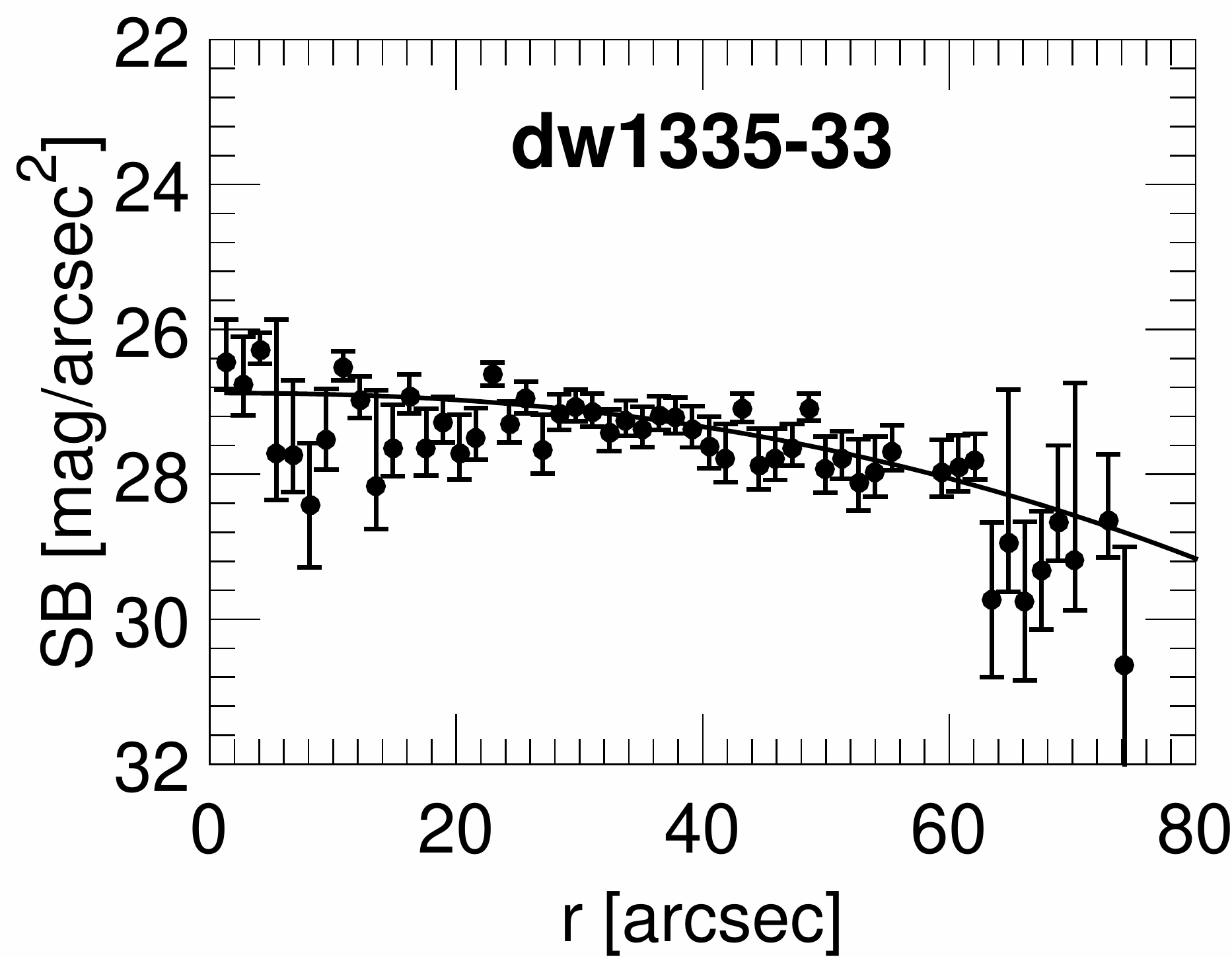}
\includegraphics[width=4cm]{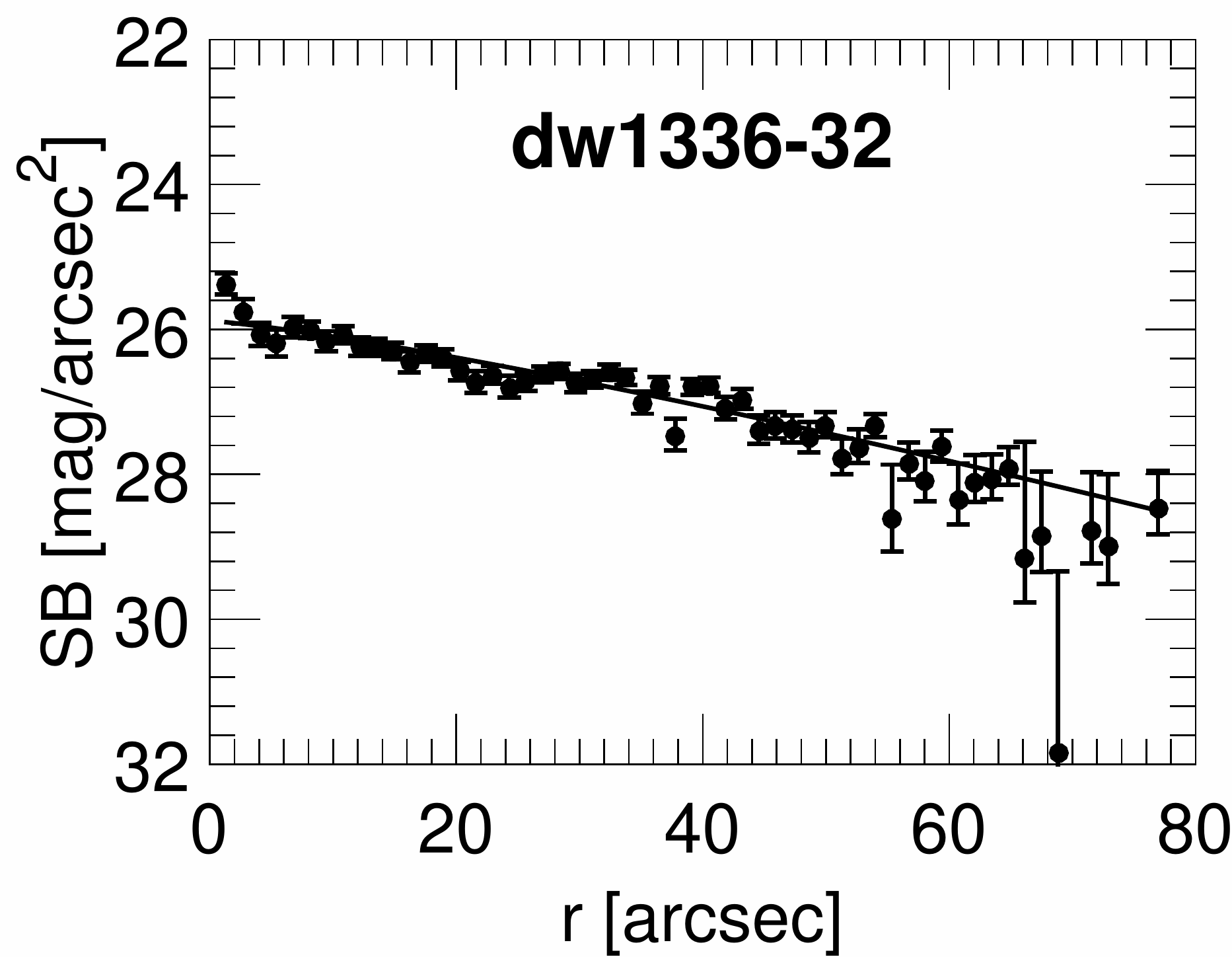} \\
\includegraphics[width=4cm]{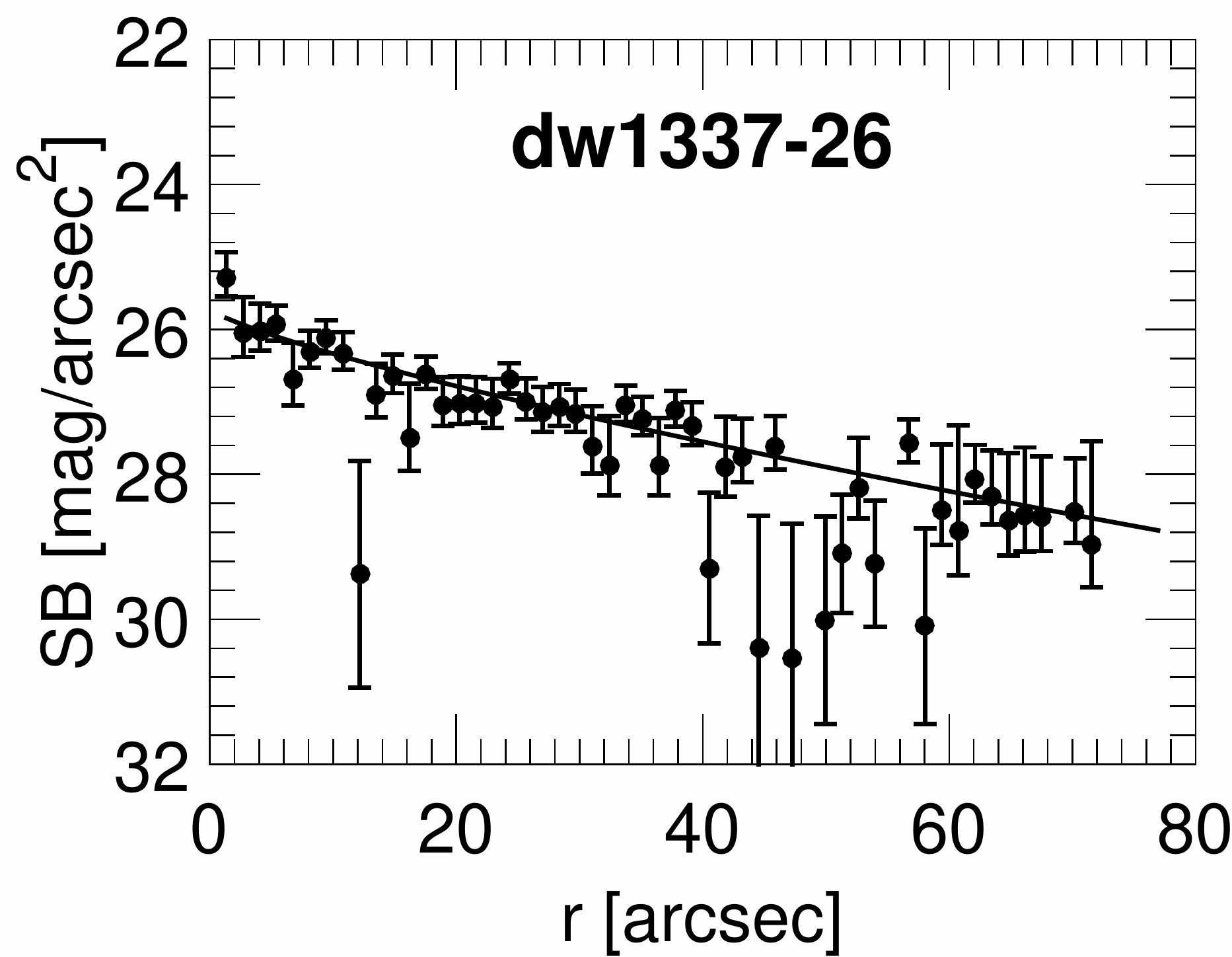}
\includegraphics[width=4cm]{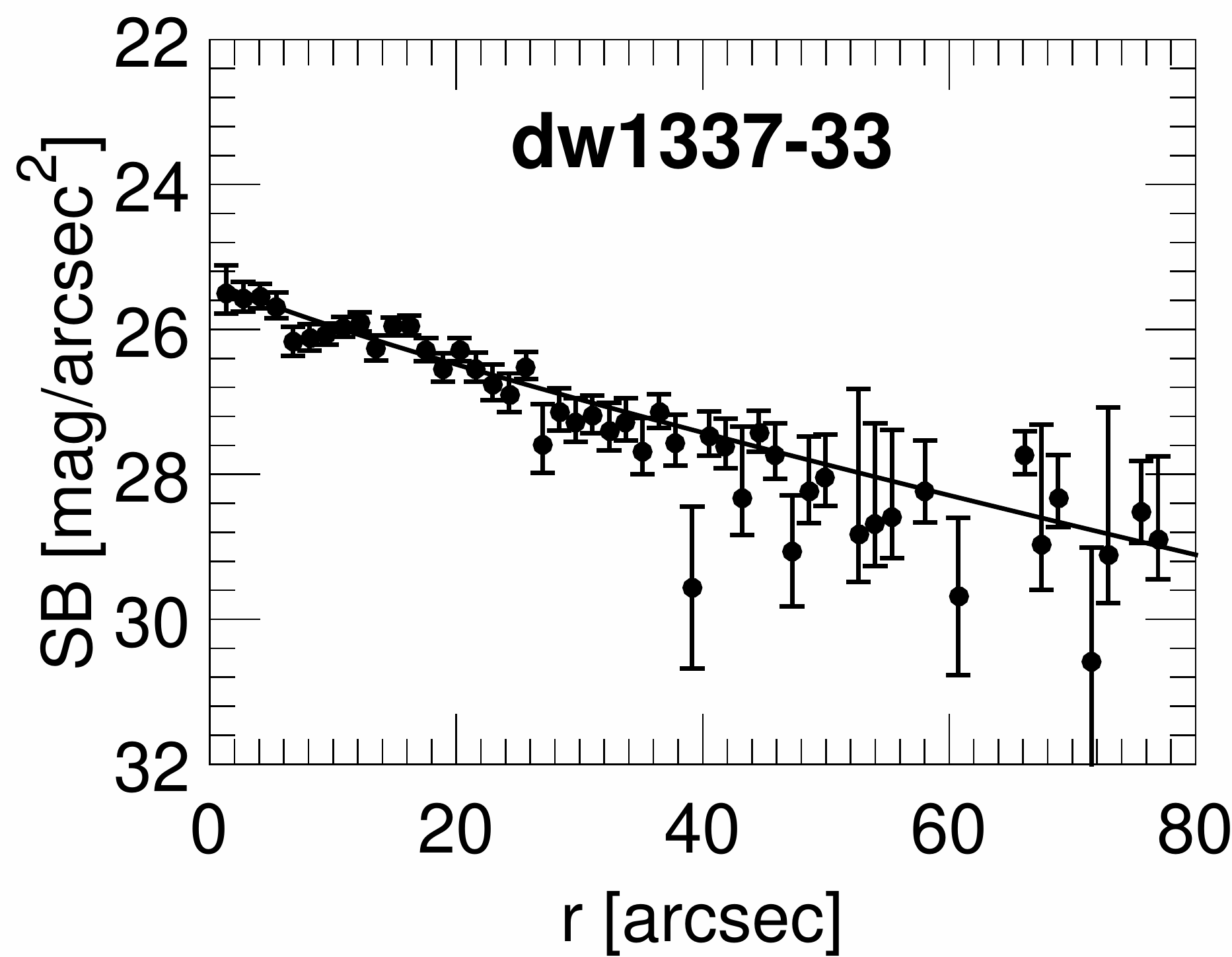}
\includegraphics[width=4cm]{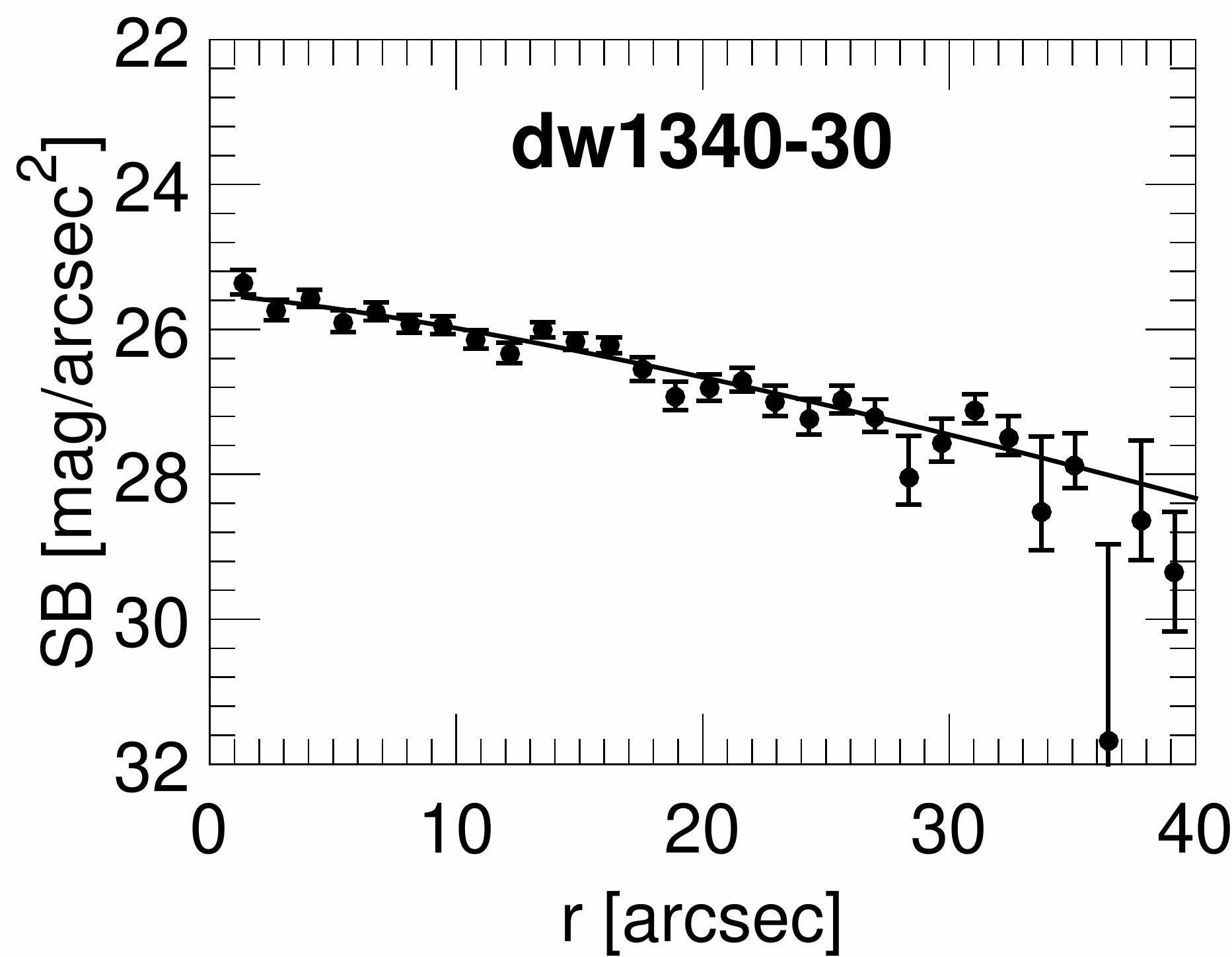}
\includegraphics[width=4cm]{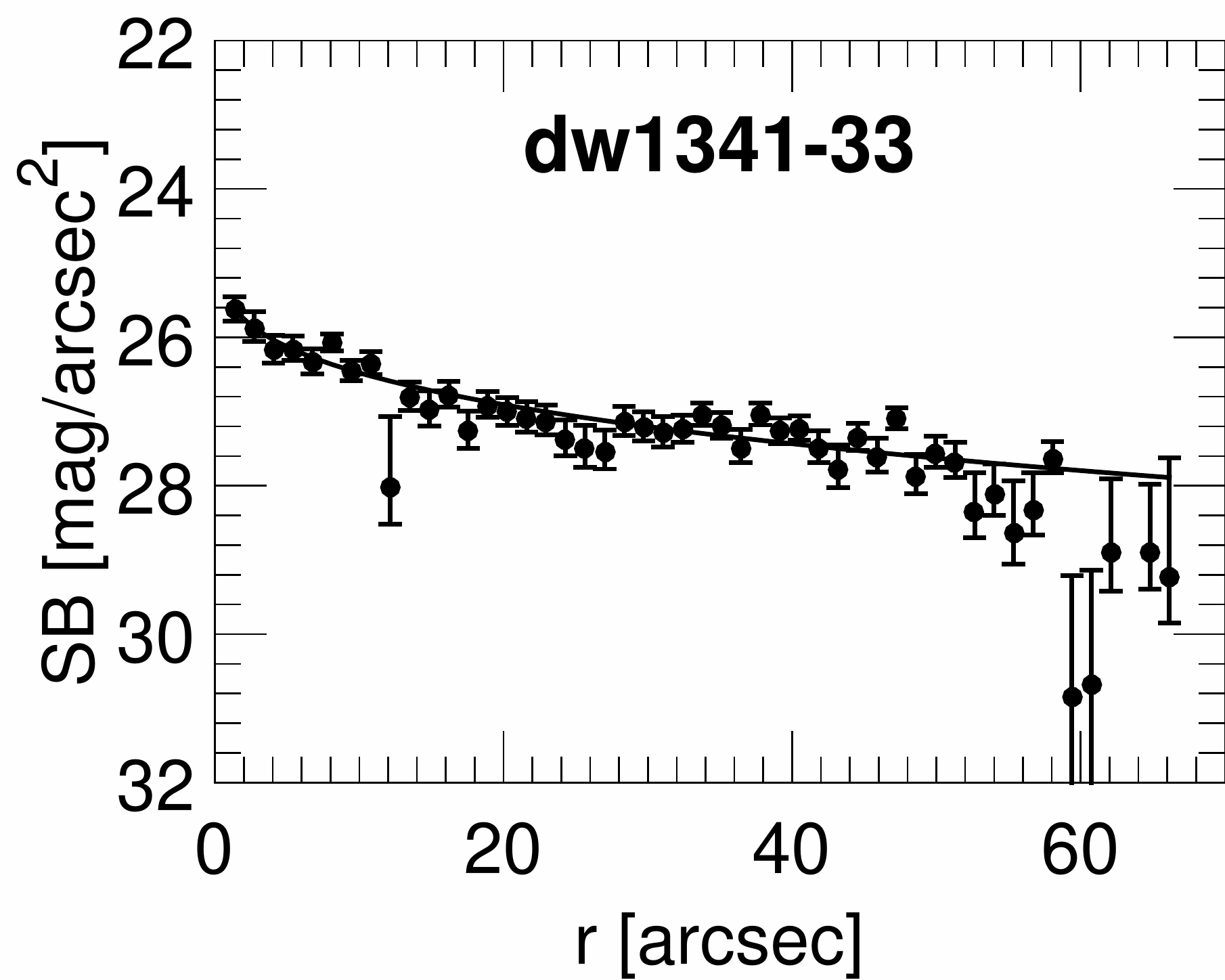} \\
\caption{Radial surface brightness profiles and best-fitting S\'ersic profiles for all dwarf candidates in the $r$ band. }
\end{figure*}

\begin{table*}
\label{table:2}
\caption{Photometric parameters of the candidates and the known dwarf galaxies in the M\,83 subgroup.}
\centering
\small
\begin{tabular}{lcccccccccccc}
\hline\hline 
Name & $m_{r}$ & $m_{g}$  & $A_{r}$ & $A_{g}$ & $M_{r}$ & $(g-r)_0$ & $\mu_{0,r}$ & $r_{0,r}$ & $n_r$ & $\langle\mu\rangle_{eff,r}$ &  $r_{eff,r}$\\ 
 & mag & mag & mag & mag & mag & mag &  mag arcsec$^{-2}$ &arcsec & & mag arcsec$^{-2}$ &arcsec  \\ 
 (1)& (2) & (3) & (4) & (5) & (6) & (7) &  (8) & (9) &(10) & (11) & (12) \\ 
\hline \\
dw1325-33 & 18.45 & 18.87 & 0.177 & 0.227 & -10.19 & 0.37 &$ 26.06  \pm  0.17  $& $19.65 \pm  3.24$ &$  0.89 \pm  0.11   $& 26.88&$18.64$\\
dw1326-29 & 17.85 & 18.67 & 0.142 & 0.205 & -10.75 & 0.75 &$ 25.09  \pm  0.08  $& $ 15.32 \pm  0.99 $&$  0.72 \pm  0.05   $& 25.63 & 13.67 \\ 
dw1326-35 & 18.13 & 18.80 & 0.142 & 0.205 & -10.47 & 0.61 &$ 24.31  \pm  0.12  $& $7.56 \pm  0.99 $  &$  1.02 \pm  0.09   $& 25.27 & 10.29 \\ 
dw1328-29 & 18.45 & 19.01 & 0.126 & 0.182 & -10.14 & 0.51 &$ 24.76  \pm  0.22 $ & $6.07 \pm  1.73 $  &$  1.37 \pm  0.17   $& 26.12 & 12.88 \\ 
dw1329-32 & 16.65 & 17.29 & 0.124 & 0.179 & -11.94 & 0.59 &$ 22.84  \pm  0.02  $& $8.83 \pm  0.15 $  &$ 0.71 \pm  0.02   $& 23.43 & 8.83 \\ 
dw1330-32 & 18.17 & 18.96 & 0.115 & 0.166 & -10.40 & 0.74 &$ 24.71  \pm  0.07  $& $11.28 \pm  0.63$ &$  0.63 \pm  0.05   $& 25.09 & 9.42 \\ 
dw1330-33 & 19.05 & 20.43 & 0.139 & 0.2   & -9.55  & 1.32 &$ 24.54 \pm  0.09  $      &       $7.40 \pm  0.58 $       & $ 0.76 \pm  0.08   $& 25.05 & 6.30 \\ 
dw1330-34 & 18.05 & 18.91 & 0.135 & 0.195 & -10.55 & 0.80 &$ 24.38  \pm  0.10  $& $ 8.10 \pm  0.89 $ &$  1.05 \pm  0.08   $& 25.25 & 10.72 \\ 
dw1334-32 & 18.13 & 18.27 & 0.134 & 0.194 & -10.47 & 0.08 &$ 26.97  \pm  0.48  $& $52.49 \pm 27.27 $&$  1.14 \pm  0.50   $& 28.14 & 38.32 \\ 
dw1335-29 & 17.82 & 18.87 & 0.104 & 0.151 & -10.74 & 1.00 &$ 25.85  \pm  0.09  $& $27.31 \pm  1.66$ & $ 0.56 \pm  0.05   $& 26.36 & 20.22 \\ 
dw1335-33 & 17.25 & 18.56 & 0.126 & 0.182 & -11.33 & 1.25 &$ 26.87  \pm  0.09  $& $57.74 \pm  2.97$ &  $0.44 \pm  0.07   $& 27.24 & 39.37 \\ 
dw1336-32 & 16.81 & 17.89 & 0.13  & 0.188 & -11.78 & 1.03 &$ 25.88  \pm  0.07  $& $37.19 \pm  2.41$ &$  0.83 \pm  0.07   $& 26.53 & 34.62 \\ 
dw1337-26 & 17.34 & 18.23 & 0.141 & 0.204 & -11.26 & 0.82 &$ 25.34  \pm  0.35  $& $11.78 \pm  6.49$ & $ 1.69 \pm  0.32   $& 27.07 & 34.61 \\ 
dw1337-33 & 17.20 & 17.82 & 0.114 & 0.165 & -11.37 & 0.56 &$ 25.30  \pm  0.16  $& $17.99 \pm  3.44$ &  $1.19 \pm  0.13   $& 26.30 & 25.51 \\ 
dw1340-30 & 17.86 & 18.70 & 0.132 & 0.191 & -10.73 & 0.78 &$ 25.50  \pm  0.10  $& $18.96 \pm  1.81$ &$  0.81 \pm  0.10   $& 26.03 & 16.86 \\ 
dw1341-33 & 17.20 & 18.22 & 0.11  & 0.159 & -11.37 & 0.97 & $25.05  \pm  0.49  $& $ 4.88 \pm  5.64 $& $ 2.70 \pm  0.62   $& 27.00 & 35.43 \\ 
\\
KK195           & 16.63 & 16.82 & 0.14 & 0.20   & -12.10 & 0.12 &$ 24.01  \pm  0.07  $& $13.33 \pm  0.84$ &  $0.88 \pm  0.05  $      & 24.79 & 16.36 \\ 
CenA-dE2  & 16.93 & 17.74 & 0.15 & 0.22         & -11.68 & 0.74 &$ 24.51  \pm  0.06  $& $17.78 \pm  0.83 $&$  0.69 \pm  0.04  $      & 25.05 & 16.34 \\ 
Cen8                    & 16.40 & 17.16 & 0.16 & 0.23   & -12.22 & 0.70 &$ 23.22  \pm  0.03 $& $11.56 \pm  0.35$& $ 0.74 \pm  0.03  $ & 23.85 & 12.04 \\ 
KK200           & 14.66 & 15.44 & 0.16 & 0.23   & -13.82 & 0.70 &$ 22.00  \pm  0.03  $& $10.21 \pm  0.29 $&  $1.15 \pm  0.01  $ & 23.32 & 21.05 \\ 
IC4247          & 13.46 & 14.05 & 0.15 & 0.21   & -15.17 & 0.52 &$ 20.14  \pm  0.02  $&$  7.62 \pm  0.14 $&$  1,12 \pm  0.01   $& 21.44 & 15.32 \\ 
ESO444-78 & 14.47 & 15.09 & 0.12 & 0.17         & -14.25 & 0.56 &$ 22.11  \pm  0.01  $&$ 15.70 \pm  0.24 $& $ 0.83 \pm  0.02  $      & 22.92 & 19.10 \\ 
HI J1337-33 & 16.81 & 17.28 & .11 & 0.16 & -11.50 & 0.42 &$ 21.76  \pm  0.03  $&$  4.36 \pm  0.09 $& $ 0.85 \pm  0.01$& 22.70 & 5.77 \\ 
ESO444-84 & 14.25 & 14.71 & 0.16 & 0.23         & -14.23 & 0.39 &$ 22.47  \pm  0.02  $&$ 22.59 \pm  0.24 $& $ 0.65 \pm  0.01 $       & 22.98 & 21.39 \\ 
NGC5253         & 9.85 & 10.36 & 0.13 & 0.18            & -18.03 & 0.45 &$ 16.88  \pm  0.02  $& $ 6.11 \pm  0.13 $&  $1.79 \pm  0.04 $  & 19.24 & 29.15 \\ 
IC4316          & 12.90 & 13.64 & 0.13 & 0.18   & -15.44 & 0.69 &$ 20.99  \pm  0.02  $&$ 12.64 \pm  0.29 $&  $1.30 \pm  0.02 $       & 22.54 & 33.24 \\ 
NGC5264         & 11.56 & 12.27 & 0.12 & 0.17   & -16.96 & 0.66 &$ 20.54  \pm  0.01  $& $27.34 \pm  0.24 $&$  0.92 \pm  0.01 $  & 21.40 & 36.44 \\ 
CenA-dE4        & 16.25 & 17.01 & 0.14 & 0.20   & -12.35 & 0.69 &$  23.79  \pm  0.04  $&$ 14.67 \pm  0.54 $& $ 0.91 \pm  0.03 $       & 24.71 & 19.27 \\ 
\hline
\end{tabular}
\end{table*}


\section{Discussion}
\begin{figure*}
\centering
  \includegraphics[width=8.5cm]{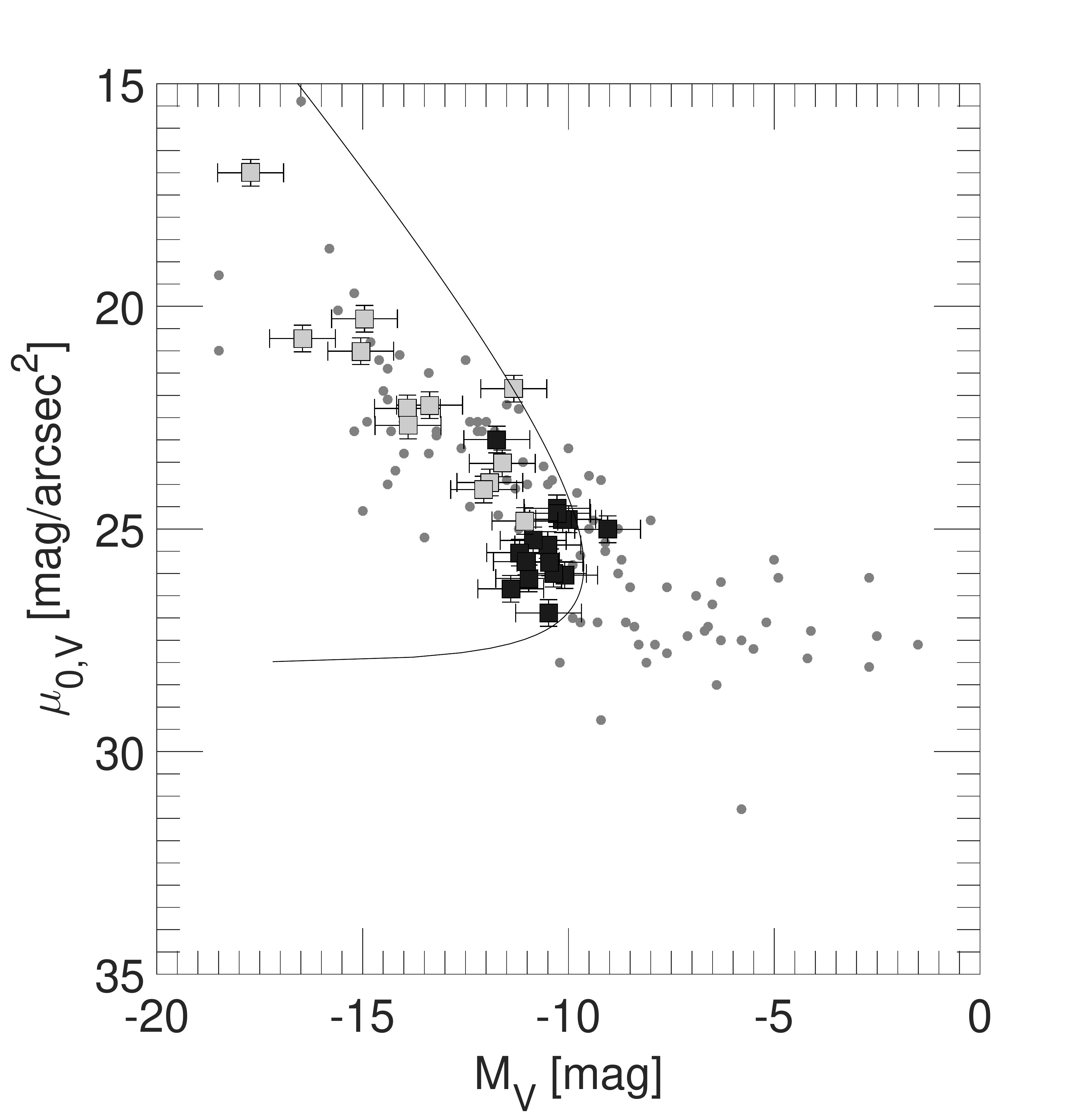}
        \includegraphics[width=8.5cm]{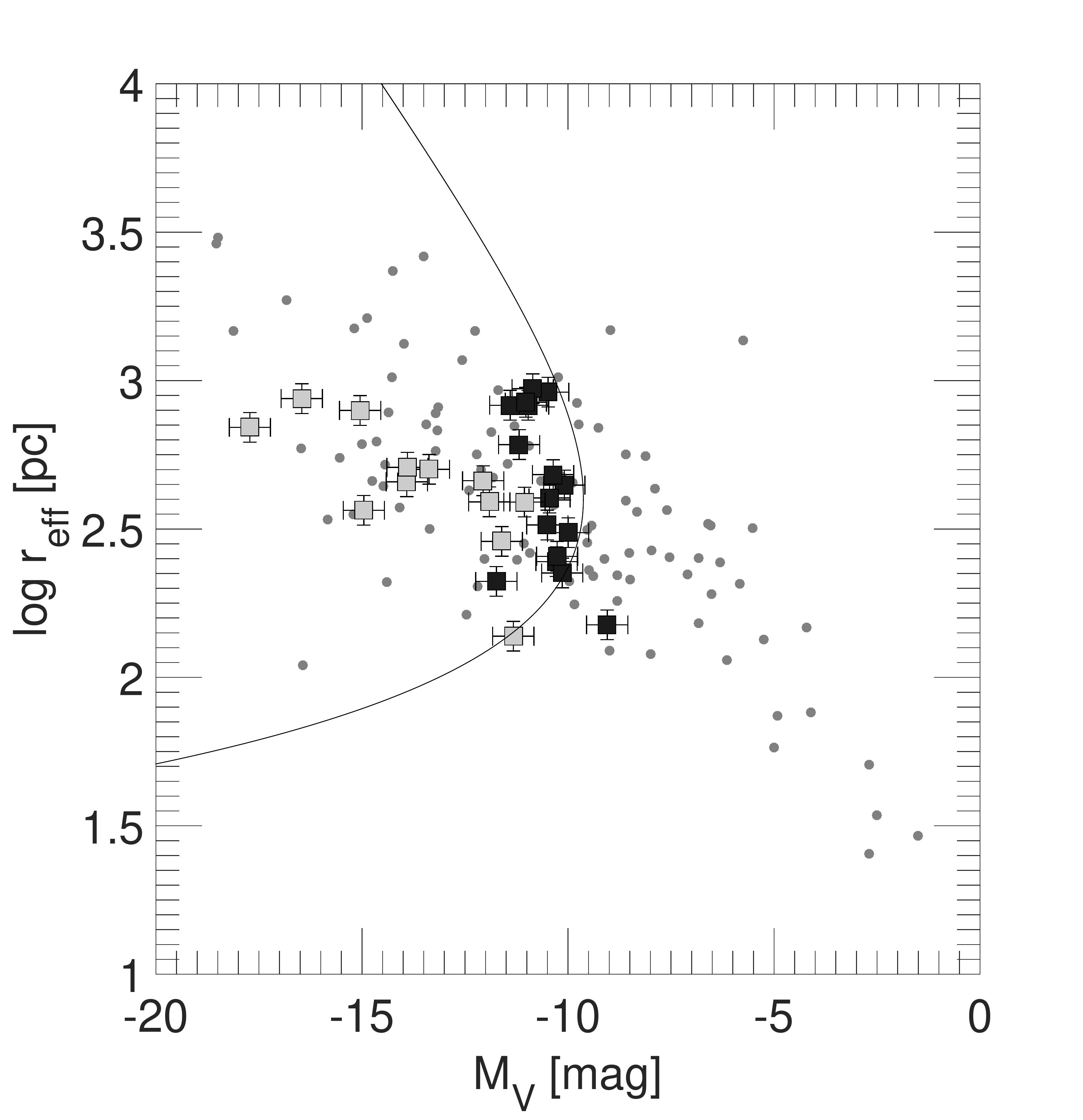}
  \caption{Left: The $\mu_0-M$ relation in the $V$ band (see text for conversion formula) for our candidates (black squares), the known dwarfs of the M\,83 subgroup (gray filled squares) and LG dwarfs \citep[dots,\,][]{2012AJ....144....4M}. { Right: The same for the effective radius, $\log r_{eff}$, instead of central surface brightness. Conservative global error bars of 0.5 mag for $M$, 0.3 mag for $\mu$, and 0.05 for $\log r$ are plotted. Absolute magnitudes assume a distance of 4.9 Mpc (M\,83).} The curves indicate the completeness limit of the present survey on the assumption that every candidate larger than 40 arcsecs in diameter within an isophotal level of 28 mag arcsec$^{-2}$ is included (see text).}
\end{figure*}

To test the credibility of our new dwarf candidates, we compared them with the photometric properties of the known dwarfs in the survey region and with the dwarfs in the Local Group. The classical tool for such a comparison is the $\mu-M$ (surface brightness -- absolute magnitude), { or equivalently the $\log r-M$ (radius -- absolute magnitude)} relation. In Fig.\,6 we have plotted the central $\mu$ (left panel) and the { effective radius, $\log r_{eff}$,} (right panel) versus $M$ in the $V$ band, which is the photometric band provided for LG dwarfs by \citet{2012AJ....144....4M}. The transformation of our $gr$ photometry into $V$ for the purpose of comparison is given above in Sect.\,2. 
Figure\,6 clearly suggests that our candidates, when placed at a nominal distance of 4.9\,Mpc, appear as a natural extension of the known members of the M\,83 subgroup toward fainter luminosities. Moreover, most candidates are in good accord with the { photometric} relations traced out by the local dwarfs -- as expected if they are indeed members of the Centaurus group.

{ We also plot here the completeness boundary curve we have  introduced in Sect.\,3, in the context of artificial galaxies. The boundary assumes that we detected all dwarf candidates that are larger than 40 arcsecs in diameter (= 2 $r_{lim}$) within an isophotal level of $\mu_{lim}$ = 28 $V$ mag arcsec$^{-2}$. Assuming exponential profiles (S\'ersic $n$ = 1), the curve in the $\mu-M$ plane (Fig.\,6 left) is given in Sect.\,3. In the $\log r_{eff}\arcsec-m$ representation the curve has the following form: 
$$m_{tot}=\mu_{lim} - \frac{r_{lim}}{0.5487 r_{eff}} - 2.5\log[2\pi\cdot (0.5958\cdot r_{eff})^2]$$
 \citep[see][]{1990PhDT.........1F, 1988AJ.....96.1520F}. To obtain the $\log r_{eff}$[pc]$-M$ relation as shown in the right
panel of Fig.\,6, the distance of 4.92 Mpc (M\,83) has to be taken into account.} 

Our candidates lie on the left side of theses boundaries, as expected, with the exception of dw1330-33, the faintest candidate in terms of total brightness, which is fairly small for its surface brightness; it is accordingly noted as a possible background object in Table 1. The one high surface brightness candidate in Fig.\,6 is the BCD candidate dw1329-32. The two cirrus candidates, dw1334-32 and dw1335-33, are at the lower surface brightness end of the completeness boundaries. Given the large cosmic scatter of the $\mu-M$ and $\log r-M$ relations, their dwarf membership cannot be excluded at this point, however.
\begin{figure}
  \resizebox{\hsize}{!}{\includegraphics{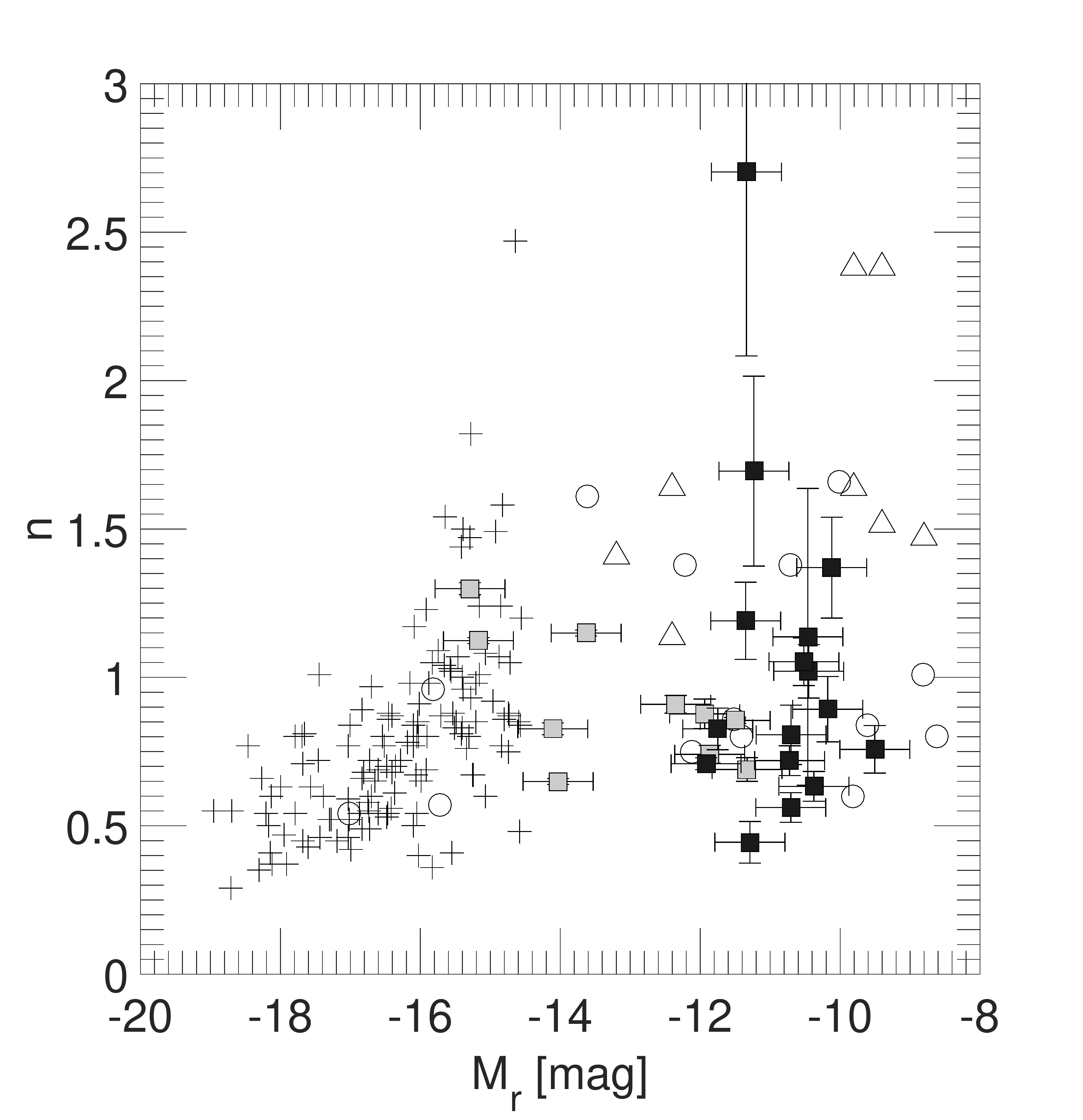}}
  \caption{S\'ersic $n-M$  relation in the $r$ band for our new dwarf candidates (black squares), the known dwarfs in the survey region (gray filled squares), Virgo dEs \citep[plus signs, ][]{1998A&A...333...17B}, LG dwarfs \citep[open circles, ][]{2000AJ....119..593J}, and M81 dwarfs \citep[open triangles, ][]{2009AJ....137.3009C}.{ The vertical error bars shown for our dwarf candidates and previously known M\,83 subgroup members are taken from Table 2. Horizontal error bars are a global $\pm$0.5 mag for their absolute magnitudes.}}
\end{figure}

In Fig.\,7 we compare our candidates with known dwarfs in the S\'ersic $n-M$ relation. The comparison sample consists of the known M\,83 subgroup dwarfs, LG dwarfs \citep{2000AJ....119..593J}, Virgo dwarfs \citep{1998A&A...333...17B}, and M\,81 group dwarfs \citep{2009AJ....137.3009C}. While at the bright end the relation is quite narrow, it spreads with decreasing luminosity.{ The large scatter is clearly not explained by photometric errors alone}. Our candidates fit well into this spread defined by the LG and confirmed M\,81 group dwarfs. The one outlier at the top is dw1341-33. With $n=2.70$ this candidate is rather cuspy (see also Fig.\,5
). This object has many foreground stars and a very diffuse shape. Improper star removal and a poor choice of the center for the radial profile could lead to an error in the best fit and therefore in the S\'ersic $n$. But in summary, both the $\mu - M$ and $n-M$ relations clearly render most of our candidates quite plausible as dwarf members of the Centaurus complex. 

Is the number of new candidates consistent with what we expect based on the suspected faint-end luminosity function of the group? Figure\,6 shows that our survey has pushed down the completeness limit by roughly 2 magnitudes, from $M_V \approx -12$ to $M_V \approx -10$. If we take the luminosity functions of other nearby groups (Cen A, M81, and And) given in Fig.\,35 of \citet{2009AJ....137.3009C} as a measure, we expect an increase of the number of group members by a factor of between 1.48 and 1.74, depending on the steepness of the faint end of the LF. For the surveyed region with 14 known members (Fig.\,1) we therefore predict (in retrospect) a discovery of between 7 and 10 new members in the given luminosity interval,{ not accounting for a detection efficiency of less than 100\%}. With actually 16 new candidates we thus have discovered even more than expected! This excess can be interpreted in different ways. The most obvious interpretation is that some of our candidates are no dwarf galaxies, after all. We have indeed singled out four doubtful cases. Another, or additional, possibility is that many of these new objects actually belong to the richer Cen\,A subgroup with which the M\,83 subgroup heavily overlaps in the sky. It is indeed very intriguing that most of the new candidates and many of the known dwarf members lie in the SW corner of the search region -- precisely in the direction to Cen\,A.{ Our artificial galaxy tests exclude the possibility that this is due to an inhomogeneous detection efficiency.} In Fig.\,1 we have drawn the direction to Cen\,A and also an estimate of the virial radius of the M\,83 subgroup to indicate its gravitational size. The virial radius is conventionally defined as the radius within which the mean mass density of a group is 200 times the critical density of the Universe, encompassing the sphere of matter that has collapsed to a halo at the present epoch \citep[e.g., ][]{2015AJ....149...54T}. For the M\,83 subgroup this amounts to 0.21\,Mpc, corresponding to 2.4 degrees in the sky. The virial radius of the more massive Cen\,A subgroup is larger with 4 degrees, but does not reach M\,83 \citep[see also ][]{2015ApJ...802L..25T}. Many, if not most of the new dwarf candidates clearly lie outside the virial radius of the M\,83 subgroup and thus might belong to the Cen\,A outskirts or a common envelope of the two subgroups.

\section{Conclusions}
We have carried out a search for new dwarf galaxies in deep images taken with the Dark Energy Camera at CTIO in an area of 60 square degrees covering the M\,83 subgroup of the Centaurus complex. We found 16 new dwarf candidates in the magnitude range 17 $< m_r <$ 19, all of which, with the exception of one BCD candidate, being of very low surface brightness, in the range 25 $< \langle \mu \rangle_{eff,r} <$ 28. 
A comparison with the photometric properties of the known brighter M\,83 subgroup members and all Local Group dwarfs in the $\mu-M$, $\log r-M$ and S\'ersic $n-M$ planes suggests that most of the new candidates are very likely dwarf members of the M\,83 subgroup, extending the LF of its known population down to $M_{r}\approx -10$. The distribution of the candidates is significantly prolonged toward the richer and somewhat closer Cen\,A subgroup. Some of the new candidates might belong to the Cen\,A subgroup or be part of a wider envelope of the Centaurus complex; these would intrinsically be somewhat brighter. Very recently,   
\citet{2015ApJ...802L..25T} reported evidence that the Cen\,A subgroup itself has a double-planar structure. Most Cen\,A members seem to lie in either of two parallel thick sheets that are separated by about 300 kpc. The M\,83 subgroup is roughly lying in the extension of the sheet going through Cen\,A. It will therefore be highly rewarding to widen and deepen the present survey and to conduct follow-up observations of the candidates, not only to confirm their Centaurus membership, but to derive TRGB distances to allow a more detailed study of the 3D structure of the Centaurus complex. 


\begin{acknowledgements}

OM and BB are grateful to the Swiss National Science Foundation for financial support. 
HJ acknowledges the support of the Australian Research Council through Discovery project DP150100862.
This project used data obtained with the Dark Energy Camera (DECam), which was constructed by the Dark Energy Survey (DES) 
collaborating institutions: Argonne National Lab, University of California Santa Cruz, University of Cambridge, Centro de Investigaciones Energeticas, Medioambientales y Tecnologicas-Madrid, University of Chicago, University College London, DES-Brazil consortium, University of Edinburgh, ETH-Zurich, Fermi National Accelerator Laboratory, University of Illinois at Urbana-Champaign, Institut de Ciencies de l'Espai, Institut de Fisica d'Altes Energies, Lawrence Berkeley National Lab, Ludwig-Maximilians Universitat, University of Michigan, National Optical Astronomy Observatory, University of Nottingham, Ohio State University, University of Pennsylvania, University of Portsmouth, SLAC National Lab, Stanford University, University of Sussex, and Texas A\&M University. Funding for DES, including DECam, has been provided by the U.S. Department of Energy, National Science Foundation, Ministry of Education and Science (Spain), Science and Technology Facilities Council (UK), Higher Education Funding Council (England), National Center for Supercomputing Applications, Kavli Institute for Cosmological Physics, Financiadora de Estudos e Projetos, Fundao Carlos Chagas Filho de Amparo a Pesquisa, Conselho Nacional de Desenvolvimento Cientfico e Tecnolgico and the Ministrio da Cincia e Tecnologia (Brazil), the German Research Foundation-sponsored cluster of excellence ``Origin and Structure of the Universe" and the DES collaborating institutions. 
\end{acknowledgements}

\bibliographystyle{aa}
\bibliography{bibliographie}




\end{document}